\documentclass{aastex62}

\usepackage[utf8]{inputenc}
\usepackage[version=4]{mhchem} 
\usepackage{amsmath}
\usepackage{amssymb}
\usepackage{graphics}
\usepackage{rotating}
\usepackage{color}
\usepackage{epstopdf}
\definecolor{red}{rgb}{1.0, 0.0, 0.0}
\definecolor{green}{rgb}{0.3, 0.73, 0.09}
\definecolor{blue}{rgb}{0.0, 0.0, 1.0}

\usepackage{ulem}

\begin{document}

\title{Ice Age : Chemo-dynamical modeling of Cha-MMS1 to predict new solid-phase species for detection with JWST}

\correspondingauthor{Mihwa Jin}
\email{mj2ua@virginia.edu}

\author{Mihwa Jin}
\affil{Department of Chemistry, University of Virginia, 409 McCormick Rd, Charlottesville, VA, 22904, USA}
\nocollaboration

\author{Ka Ho Lam}
\affiliation{Department of Astronomy, University of Virginia, 530 McCormick Rd, Charlottesville, VA, 22904, USA}
\nocollaboration

\author{Melissa K. McClure}
\affiliation{Leiden Observatory, Leiden University, PO Box 9513, 2300 RA Leiden, The Netherlands}
\nocollaboration

\author{Jeroen Terwisscha van Scheltinga}
\affiliation{Leiden Observatory, Leiden University, PO Box 9513, 2300 RA Leiden, The Netherlands}
\affiliation{Laboratory for Astrophysics, Leiden Observatory, Leiden University, PO Box 9513, 2300 RA Leiden, The Netherlands}
\nocollaboration

\author{Zhi-Yun Li}
\affiliation{Department of Astronomy, University of Virginia, 530 McCormick Rd, Charlottesville, VA, 22904, USA}
\nocollaboration

\author{Adwin Boogert}
\affiliation{Institute for Astronomy, University of Hawaii, 2680 Woodlawn Drive, Honolulu, HI, 96822, USA}
\nocollaboration

\author{Eric Herbst}
\affiliation{Department of Chemistry, University of Virginia, 409 McCormick Rd, Charlottesville, VA, 22904, USA}
\affiliation{Department of Astronomy, University of Virginia, 530 McCormick Rd, Charlottesville, VA, 22904, USA}
\nocollaboration

\author{Shane W. Davis}
\affiliation{Department of Astronomy, University of Virginia, 530 McCormick Rd, Charlottesville, VA, 22904, USA}
\nocollaboration

\author{Robin T. Garrod}
\affiliation{Department of Chemistry, University of Virginia, 409 McCormick Rd, Charlottesville, VA, 22904, USA}
\affiliation{Department of Astronomy, University of Virginia, 530 McCormick Rd, Charlottesville, VA, 22904, USA}
\nocollaboration

\begin{abstract}

Chemical models and experiments indicate that interstellar dust grains and their ice mantles play an important role in the production of complex organic molecules (COMs). To date, the most complex solid-phase molecule detected with certainty in the ISM is methanol, but the James Webb Space Telescope (JWST) may be able to identify still larger organic species. In this study, we use a coupled chemo-dynamical model to predict new candidate species for JWST detection toward the young star-forming core Cha-MMS1, combining the gas-grain chemical kinetic code \textit{MAGICKAL} with a 1-D radiative hydrodynamics simulation using \textit{Athena++}. With this model, the relative abundances of the main ice constituents with respect to water toward the core center match well with typical observational values, providing a firm basis to explore the ice chemistry. Six oxygen-bearing COMs (ethanol, dimethyl ether, acetaldehyde, methyl formate, methoxy methanol, and acetic acid), as well as formic acid, show abundances as high as, or exceeding, 0.01\% with respect to water ice. Based on the modeled ice composition, the infrared spectrum is synthesized to diagnose the detectability of the new ice species. The contribution of COMs to IR absorption bands is minor compared to the main ice constituents, and the identification of COM ice toward the core center of Cha-MMS1 with the JWST NIRCAM/Wide Field Slitless Spectroscopy (2.4-5.0 $\micron$) may be unlikely. However, MIRI observations (5-28 $\micron$) toward COM-rich environments where solid-phase COM abundances exceed 1\% with respect to the water ice column density might reveal the distinctive ice features of COMs.

\end{abstract}


\section{Introduction} \label{sec:intro}
Over the past decades, the rich chemistry of the interstellar medium (ISM) has been unveiled largely through dedicated ground-based microwave and sub-mm observations of molecular rotational transitions. Complex organic molecules (COMs), typically defined as carbon-bearing molecules containing 6 or more atoms, are found to be ubiquitous in the gas throughout various evolutionary stages of star formation, from extremely cold prestellar cores to chemically-rich sources such as hot corinos~\citep[e.g.][]{blake87, arce08, bottinelli10, oberg10, bacmann12, jorgensen12, fayolle15, jorgensen16}.

Chemistry taking place both in the gas and on the surfaces of interstellar dust grains has been proposed to explain such a high degree of chemical complexity in star-forming regions. Earlier theories relied more or less exclusively on gas-phase mechanisms such as ion-molecule reactions to drive the chemical complexity~\citep[e.g.][]{hamberg10, vasyuninandherbst13}, following the sublimation of simple ice species from the grains. However, the timescales available for gas-phase COM production have been argued to be too short \citep[e.g.][]{aikawa08}, encouraging the deeper exploration of an alternative explanation -- grain-surface/ice chemistry. In this scenario, COMs would already have been formed on the grains or within their ice mantles prior to ice sublimation and subsequent detection in the gas-phase. Although there has recently been renewed interest in gas-phase processes, including radiative association and neutral-neutral reactions \citep{balucani15, shannon13}, grain-surface chemistry is nevertheless considered to be -- at minimum -- a substantial contributor to COM production in star-forming regions. The recent modeling study of hot core chemistry by \citet{garrod21} indicates that in those hot sources, at least, COM production on grains outweighs that occurring in the gas phase.

Until recently, the grain-surface/ice chemistry was believed to be related to the warming of the dust, with intermediate temperatures ($\textrm{T}_\textrm{dust} \ga$ 30 K) being sufficient to allow heavy radicals on the grains to become thermally mobile and thus reactive \citep{garrodandherbst06}. However, recent detection of COMs toward prestellar cores shows that large species can be synthesized during the very cold, early phase of star formation in which thermal diffusion of radicals is highly inefficient. Laboratory studies, and now gas-grain chemical models, show that the formation of COMs appears to be possible in such cold and quiescent environments through nondiffusive reactions, and/or radiolysis within ice mantles~\citep{fedoseev15, chuang2016, shingledecker18, garrod19, jinandgarrod20}. This new picture of grain-surface/ice chemistry during star formation suggests that much of the gas-phase COM material observed under hot conditions may have been formed during the very earliest stages of ice-mantle growth \citep{garrod21}, although some processing likely occurs at later times and higher temperatures. A recent laboratory study shows that species as complex as glycine (NH$_2$CH$_2$COOH) may also be formed under cold, prestellar conditions, given an appropriate ice composition \citep{Ioppolo20}.

Although not directly detected, the presence of highly complex molecules on grain surfaces has frequently been inferred, based on a variety of evidence. For example, many laboratory studies indicated that UV and energetic radiation of ices results in the formation of COMs~\citep[e.g.][]{gerakines96, bernstein02}. The chemical richness revealed in hot cores by gas-phase observations may indirectly indicate the formation of COMs in interstellar ices; in this scenario, an embedded protostar heats up the surrounding envelope, the ensuing thermal release of simple solid-phase molecules would be accompanied by the release of complex species that are also present in the ice~\citep[see the review of][]{jorgensen20}.

However, while the inventory of gas-phase COMs has continued to expand, our understanding on ice-mantle composition in interstellar clouds and star-forming sources has been limited to the most abundant, simple ice species ($\ce{H2O}$, $\ce{CO}$, $\ce{CO2}$, $\ce{NH3}$, $\ce{CH4}$, $\ce{H2CO}$, and $\ce{CH3OH}$) in absorption from 3 to 15 $\micron$. Several absorption features have been attributed to frozen COMs, although the identification with particular molecular species remains a hotly debated topic, in contrast to gas phase studies~\citep[reviewed by][]{boogert15}. The observational identification of larger ice species is challenging; in addition to the lack of infrared (IR) spectroscopic data from the laboratory, the spectral resolution and sensitivity of past astronomical instruments were insufficient to identify the weak and complicated IR bands of COMs, and to distinguish them from the deep, broad features presented by the abundant simple molecules. Furthermore, the detection of various ice species \citep[see Figure 7 of][]{boogert15} requires a line of sight of sufficiently high visual extinction, thus narrowing the number of sources in which COMs could plausibly be detected.

With the launch of the James Webb Space Telescope (JWST), our limited observational knowledge of solid-phase COMs is expected to expand rapidly, particularly through the approved JWST Early Release Science (ERS) program {\em Ice Age}. This program will trace the evolution of pristine and complex ice chemistry during different evolutionary stages of low-mass star formation, from prestellar cores to protoplanetary disks, through observations of the Ced 110 region of the Chamaeleon molecular cloud I (Cha I) complex, located at a distance of 192 pc~\citep{dzib18}. Many background stars exist within this field, providing more than 140 lines of sight, which would allow the determination of spatial profiles for ice abundances. In particular, four of these lines of sight probe the ice within 10,000 AU of the Class 0 protostar ``Chamaeleon-MMS1" (Cha-MMS1) which is a main target of this study. Due to the benefits of observing ice formation in this field, one of the main science goals of the {\em Ice Age} project is to identify new COM species within ice mantles. If successful, such identifications would provide a direct observational link between the well-known gas-phase presence of COMs in warm/hot sources and their expected production on dust grains at earlier times.

In preparation for the {\em Ice Age} project, we have modeled the time-dependent chemical evolution and structure of one of its target sources, Cha-MMS1, to guide the observational study of solid
 state COMs towards this object. This very young star-forming core is an interesting source from an astrophysical perspective. Since \citet{reipurth96} first identified this source, it has drawn attention as a promising candidate for the first hydrostatic core (FHSC) stage of evolution due to its weak mid-IR emission at 24$\micron$ and 70$\micron$~\citep{belloche06}. The FHSC is defined as an adiabatic core in a quasi-hydrostatic equilibrium state where the gravity is balanced by the thermal pressure gradient~\citep{larson69}. Because of the short lifetime of this stage, only one FHSC candidate, Barnard 1b-N~\citep{gerin15}, has so far been successfully identified as such. High-resolution observations toward Cha-MMS1 have recently become available with ALMA, but the evolutionary stage of this source remains highly debated. \citet{busch20} detected a CO outflow towards Cha-MMS1 with high angular resolution observations. The deprojected outflow velocities exceed the range of the expected values from a FHSC outflow, implying that Cha-MMS1 has already gone through the FHSC phase; those authors suggest a dynamical age of 200--3000 yr post-FHSC. However, another recent paper, \citet{maureira20}, reported weak or missing emission of $\ce{HCO+}$, HCN, and $\ce{CH3OH}$ toward the Cha-MMS1 envelope. ALMA observations of $\ce{CO}$, $\ce{CS}$, $\ce{H2CO}$, and $\ce{CH3OH}$ towards an outflow associated with Cha-MMS1 also show that the outflow is weaker than the ones produced by Class 0 objects~\citep{allen20}. However, regardless of whether Cha-MMS1 is in the FHSC stage or is a young Class 0 source, we would expect the state of the ice mantles, especially in the more extended regions, to be largely unaffected by this change.

In this work, the chemical structure of Cha-MMS1 is modeled by using an astrochemical model combined with a radiative hydrodynamics simulation (RHD) of FHSC with a simple 1-D approach. This paper is structured as follows: The chemical model and its implementation within a dynamical simulation are described in \S~2. The results of the models are discussed in \S~3. Conclusions are summarized in \S~4.

\section{Chemo-dynamical model} \label{sec:model}

The main modeling focus of this study involves the coupling of a three-phase chemical kinetics model to the outputs of a radiation hydrodynamics simulation (RHD) of the formation of the first hydrostatic core, in order to investigate the chemistry of Cha-MMS1. This chemo-dynamical modeling method enables us to explore the chemical responses to the dynamical change, achieving a more realistic chemical picture of the source. Thus far, several models have employed this approach to understand the dynamical / chemical evolution of FHSC. For example, \citet{furuya12} performed three-dimensional RHD simulation coupled with astrochemical models and showed that large organic molecules are associated with the FHSC. \citet{hincelin13, hincelin16} developed the same kind of simulation with the consideration of the effect of magnetic field. Although these models are powerful, the chemical treatments of these studies were focused on the gas and grain-surface chemistry only, using a so-called two-phase approach. However, full consideration of the chemistry including deeper icy mantle layers is needed particularly for the investigation of astronomical ices, because of the preservation of the chemical composition of earlier surface layers during the evolution of the cloud~\citep{garrodandpauly11}. Here, we use a fully-coupled gas-phase / grain-surface / icy grain-mantle chemical model that also includes non-diffusive surface/ice-mantle chemistry \citep{jinandgarrod20}. This allows us to build up simulated profiles of gas- and solid-phase species through the core that may ultimately be compared with new data from the JWST {\em Ice Age} project.
The approach used here is broadly similar to that of \citet{Barger21}, who applied a 1-D RHD method to the evolution of high-mass hot cores (note also that the gas-grain chemical model used by those authors included only diffusive grain-surface/ice-mantle chemistry, whereas the present model also employs non-diffusive chemistry).

Firstly, a one-dimensional collapse from a flat density profile ($n_\mathrm{H} \simeq 6.3 \times 10^4$ cm$^{-3}$) to a self-supported FHSC is simulated using the hydrodynamic code {\em Athena++}~\citep{stone20}. The model setup is very similar to that presented by \citet{masunaga98}. Lagrangian particles are placed throughout the initial density profile, which then evolve over time to track the motions of mass-conserved fluid parcels, or shells, as the core collapses. The time-dependent physical conditions and radial positions of each particle (which we refer to as trajectories) determined by the dynamical code are then fed into the chemical model, {\em MAGICKAL}, to build up a corresponding chemical picture of the core. The chemical model is run independently for each trajectory, after the RHD simulation is complete; thus, there is no direct chemical interaction between the different trajectories, and the chemistry does not affect the dynamics in any way.

A total of 47 trajectories were selected to describe the time-dependent 1-D physical structure, for each of which the chemical evolution was calculated. Each trajectory has its own set of time-dependent density, temperature, visual extinction and radius values. Due to the 1-D nature of the simulations, the radial ordering of the trajectories is maintained. The visual extinction for each trajectory throughout its physical evolution (which is needed for the chemical rates and to determine the dust temperature) is calculated in post-processing, by evaluating the one-dimensional integral of the total hydrogen density profile to determine the total H column density, then applying the relation of \citet{bohlin78}, i.e.
\begin{equation}
A_\textrm{V}=\frac{3.1}{5.8\times10^{21}}N_\textrm{H}.
\end{equation}
\noindent Here, $N_\textrm{H}$ is the column density from the instantaneous position of a particular trajectory to the outer edge of the core. To this value for $A_\textrm{V}$, an assumed background visual extinction provided by material surrounding the core, $A_\textrm{V,bac}$, is added, to account for external material that is not explicitly considered in the RHD calculations.
A value of $A_\textrm{V,bac}=2.5~\textrm{mag}$ is adopted, which corresponds to the observed visual extinction threshold of 5 mag for starless cores in the Cha I cloud~\citep{belloche11}, within which Cha MMS-1 resides.

The initial conditions of the RHD simulation are intended to represent a gravitationally-bound dense condensation within the dark cloud (slightly more massive than the Jeans mass). The chemistry of this initial state should be expected already to have evolved somewhat; thus, a prior stage of collapse is also modeled, allowing the chemistry to evolve from that appropriate to a translucent cloud to that of the dense core from which the RHD model begins. This pre-RHD chemical model (referred to hereafter as the ``pre-model'') is run individually for each RHD-model trajectory, and consists of a simple (0-D) freefall collapse treatment, in which the gas density evolves from $3 \times 10^3$ cm$^{-3}$ to the (uniform) starting density of the dense core of $\sim 6.3 \times 10^4$ cm$^{-3}$ over a period of approximately 1 Myr \citep[see e.g.][]{nejad90}. The visual extinction at each position is scaled during the pre-model evolution according to $A_\textrm{V} \propto n_\mathrm{H}^{2/3}$, in keeping with past implementations of the collapse treatment \citep[e.g.][]{garrod21}, such that the visual extinction profile that is present at the initiation of the RHD simulations is attained at the end of the pre-model. The additional background extinction is added to all calculated values, as described above.

Gas temperatures in the pre-model and main chemical model are set to 10~K or to the RHD model-calculated value, whichever is greater. The dust temperature used in the chemical models is chosen as the greater of the RHD-model value and the visual extinction-dependent value obtained from the relation provided by \citet{garrodandpauly11}. Both dust temperature treatments assume a minimum value of 8~K. The details of the RHD simulation are described in \S~\ref{subsec:rhd-model}. A more detailed description of the chemistry used in the model is provided in \S~\ref{subsec:chem-model}.

\subsection{Chemical model} \label{subsec:chem-model}
The chemical simulations conducted for each of the dynamical trajectories use the astrochemical gas-grain kinetics code {\em MAGICKAL}. The coupled gas-phase, grain-surface, and icy mantle chemistry are simulated by solving a set of rate equations~\citep{garrodandpauly11, garrod13}, supplemented by the modified rate-equation treatment presented by \citet{garrod08b}.

Following \citet{jinandgarrod20} and \citet{garrod21}, the chemical model includes both the typical diffusive grain-surface/ice chemistry and a range of nondiffusive mechanisms by which grain/ice surface or bulk-ice species may meet and thence react. These mechanisms are important to treat accurately the production of COMs at low temperatures. The model includes the following nondiffusive mechanisms: three-body (3-B) reactions, in which a preceding surface/bulk reaction is followed by the immediate reaction of the product with some pre-existing nearby species; photodissociation-induced (PDI) reactions, in which a radical produced through photodissociation of a precursor instantly meets a reaction partner in its immediate vicinity; and the Eley-Rideal process, in which gas-phase species adsorb directly onto a grain-surface reaction partner. The ``PDI2'' treatment of \citet{garrod21} is employed, in which (non-hydrogenic) photodissociation products in the bulk ice that do not find an immediate reaction partner are allowed to recombine with each other. The three-body excited-formation (3-BEF) mechanism proposed by \citet{jinandgarrod20}, in which the chemical energy released by a preceding reaction allows the energized product to overcome the activation energy barrier to the follow-on reaction, is also used, adopting the generalized treatment of \citet{garrod21}. All chemical reactions included in the surface/bulk network are able to occur through both diffusive and nondiffusive mechanisms (excluding the 3-BEF mechanism, which occurs only in cases where the initiating reaction produces sufficient energy to overcome the barrier to the subsequent reaction). A discussion of the effects of nondiffusive processes in the model is provided in \S~\ref{subsec:non-diff}.

Importantly, the present model \citep[following][]{garrod21} eliminates bulk diffusion in the ice for all species but atomic and molecular hydrogen. Although this means that thermally-driven, diffusive bulk-ice reactions between radicals do not occur, those radicals may still react through the nondiffusive mechanisms described above. H atoms may also react diffusively with those radicals, as well as with more stable species such as CO via activation energy barrier-mediated reactions. Those diffusive hydrogen atoms are typically produced via photodissociation of molecules. Thus, both the direct photodissociation-driven ``PDI'' chemistry and the indirect reactions between atomic H and other species are ultimately the result of UV-induced photodissociation of bulk-ice molecules.

The photodissociation rates of grain-surface/bulk species is assumed to be a factor 3 lower than the corresponding gas-phase processes, although the other rate coefficients and the photoproducts are the same. This is based on the modeling study of \citet{Kalvans18} that theoretically determines an average, general ratio between the photodissociation coefficients for molecules in ice and gas.

\begin{deluxetable}{ccccc}
\tablewidth{0pt}
\tabletypesize{\footnotesize}
\tablecolumns{5}
\tablecaption{Initial elemental and chemical abundances \label{init-element}}
\scriptsize
\tablehead{
\colhead{Species, \textit{i}} & \colhead{$n(i)/n_\textrm{H}^{a}$ \tablenotemark{a}
}}
\startdata
$\ce{H}$ & 5.0(-4) \\
$\ce{H2}$ & 0.49975\\
$\ce{He}$ & 0.09\\
$\ce{C}$ & 1.4(-4)\\ 
$\ce{N}$ & 2.1(-5)\\
$\ce{O}$ & 3.2(-4)\\
$\ce{S}$ & 8.0(-8)\\
$\ce{Na}$ & 2.0(-8)\\
$\ce{Mg}$ & 7.0(-9)\\
$\ce{Si}$ & 8.0(-9)\\
$\ce{P}$ & 3.0(-9)\\
$\ce{Cl}$ & 4.0(-9)\\
$\ce{Fe}$ & 3.0(-9)\\
\enddata
\tablecomments{$^{a}A(B)=A^{B}$}
\end{deluxetable}

Initial elemental abundances used in the chemical model are shown in Table~\ref{init-element}. These values are partially based on those used by \citet{wakelamandherbst08} and \citet{jinandgarrod20}. However, here we adopt the ``low-metal" value \citep{graedel82} for nitrogen. In the dark clouds that should provide the starting point for low-mass star formation, 
metals (all elements other than hydrogen and helium) are depleted in the gas compared to diffuse clouds, because heavy elements have already been incorporated into solids as the density evolves. \citet{jenkins09} suggests a relationship to describe the degree of depletion of each metal element with density, and \citet{wakelamandherbst08} define the initial elemental abundances for chemical models by applying Jenkins' relation to a typical density of dense clouds (2$\times10^4 \textrm{cm}^{-3}$). Previously, we did not use the low-metal value only for nitrogen, because nitrogen is the element that does not show distinctive depletion with density evolution~\citep{jenkins09}. Here, we assume some depletion for nitrogen in dense clouds, consistent with other elements; as argued by Jenkins, observational bias could be possibly involved in measuring the nitrogen depletion. The newly adopted initial abundance for nitrogen is around 3 times lower than the values found in the $\zeta$ Oph diffuse cloud~\citep{hincelin11}. In the models, this change has the effect of reducing the ultimate solid-phase ammonia abundance to a value in keeping with observations of interstellar ice features, ensuring that COMs whose production derives from ammonia also will not be overproduced for this reason. 

\begin{deluxetable}{clccl}
\tablewidth{0pt}
\tabletypesize{\footnotesize}
\tablecolumns{5}
\tablecaption{Updated and/or newly-included grain-surface and ice-mantle reactions \label{table-rxns_update}}
\scriptsize
\tablehead{
\colhead{Related species} & \colhead{Reaction} & \colhead{$E_\textrm{A}$ (K)} & \colhead{Width (\AA)} & \colhead{Notes}}
\startdata
Formic acid & $\ce{CO}+\ce{OH} \rightarrow \ce{CO2}$ + H / COOH & 80 & 1.0 & Branching ratio updated - 99:1 $\rightarrow$ 199:1 \\
Ethanol & $\ce{H}+\ce{C2H2} \rightarrow \ce{C2H} + \ce{H2}$ & 1300 & 1.0 & New \tablenotemark{a} / 50$\%$ branching ratio versus hydrogenation \\
     & $\ce{H}+\ce{C2H3} \rightarrow \ce{C2H2} + \ce{H2}$ & 0  & 1.0 & New / 50$\%$ branching ratio versus hydrogenation  \\%
     & $\ce{H}+\ce{C2H4} \rightarrow \ce{C2H5}$ & 1040  & 1.0 & $E_\textrm{A}$ updated - 605 K $\rightarrow$ 1040 K  \tablenotemark{b} \\
     & $\ce{H}+\ce{C2H5} \rightarrow \ce{C2H4} + \ce{H2}$ & 0 & 1.0 & New / 50 \% branching ratio versus the formation of EtOH \\
     & $\ce{O}+\ce{C2H5} \rightarrow \ce{C2H4} + \ce{OH}$ & 0  & 1.0 & New / 100$\%$ branching ratio versus hydrogenation \\
     & $\ce{OH}+\ce{C2H5} \rightarrow \ce{H2O} + \ce{C2H4}$ & 0  & 1.0 & New / 50$\%$ branching ratio versus the formation of EtOH \\
     & $\ce{H}+\ce{C2H4OH} \rightarrow \ce{C2H3OH} + \ce{H2}$ & 0 & 1.0 & New / 50$\%$ branching ratio versus hydrogenation
\enddata
\tablenotetext{a}{The same $E_\textrm{A}$ assumed for H+\ce{C2H2} $\rightarrow$ \ce{C2H3} based on \citet{baluch92}}
\tablenotetext{b}{\citet{lee78}}
\end{deluxetable}

\subsubsection{Chemical reaction updates} \label{subsubsec:rxns}

The underlying chemical network used in these models is based on that presented by \citet{garrod21}, with a number of additions and changes made to the grain-surface/bulk-ice portion of the network. Particular adjustments are made to the reaction schemes involving the formation of formic acid (HCOOH, FA) and ethanol (C$_2$H$_5$OH, EtOH), which are summarized in Table~\ref{table-rxns_update}. 

Two reactions on grain surfaces mainly contribute to the formation of FA:
\begin{subequations}
\begin{align}
\rm HCO~+~OH & \rightarrow  \rm CO + \ce{H2O} \\
\rm          & \rightarrow  \rm HCOOH
\end{align}
\end{subequations}
\begin{subequations}
\begin{align}
\rm CO~+~OH  & \rightarrow  \rm \ce{CO2}~+~H \\
\rm          & \rightarrow  \rm COOH.
\end{align}
\end{subequations}
Reaction (2) has two equally weighted branches resulting in stable products, one of which is formic acid. Reaction (3), which is mediated by a modest activation energy barrier, typically leads to carbon dioxide production, but may also produce the radical COOH, which can easily be transformed into FA through hydrogenation by mobile atomic H. The branching ratio (BR) to form COOH versus \ce{CO2} + H is not well constrained. \citet{garrod21} proposed a 1\% conversion, although the resulting gas-phase formic acid in those hot core models, derived from grain-surface production, was overproduced by a factor 2 compared with well-observed sources Sgr B2(N2) and IRAS 16293-2422 (high- and low-mass hot core/corino sources, respectively). We here adjust the ratio for COOH production downward by a factor 2, to 0.5\%.

The production of ethanol on the grains in this model is strong, and can occur early in the evolution. As a result, we paid particular attention to ensuring that the grain-surface network is as complete as possible. Two reactions mainly contribute to the formation of EtOH in our chemical model:
\begin{subequations}
\begin{align}
\rm CH_{2}~+~CH_{3}OH & \rightarrow \rm CH_{3}~+~CH_{2}OH \\
\rm                   & \rightarrow \rm C_{2}H_{5}OH
\end{align}
\end{subequations}
\begin{equation}
\rm C_{2}H_{4}OH/C_{2}H_{5}O~+~\ce{H} \rightarrow \rm C_{2}H_{5}OH
\end{equation}
The first of these involves the barrier-mediated abstraction of an H-atom from the methyl end of the methanol molecule, by methylene, CH$_2$. This is allowed either to form simply two radicals, or for those radicals to immediately recombine to produce ethanol. A branching ratio of 50\% is assumed, following the approach of \citet{garrod21}.

In reaction (5), the \ce{C2H4OH} and \ce{C2H5O} radicals are formed through ongoing hydrogenation of surface and bulk-ice hydrocarbons; \ce{C2H} -- \ce{C2H2} -- \ce{C2H3} -- \ce{C2H4} -- \ce{C2H5} -- \ce{C2H6}. However, in the previous chemical network, only addition reactions were considered, rendering the efficient hydrogenation of unsaturated hydrocarbons inevitable and providing a bias toward EtOH production. To avoid overproduction, a number of backward reactions were also included, some of which have activation energy barriers (Table~\ref{table-rxns_update}).

\subsection{Radiative-hydrodynamics simulations} \label{subsec:rhd-model}

The radiative-hydrodynamic equations are solved as described by \citet{Barger21} with an additional cosmic-ray heating term. For reference, the equations are
\begin{equation}
    \frac{\partial \rho}{\partial t} + \nabla \cdot \left( \rho \mathbf{v} \right) = 0,
\end{equation}
\begin{equation}
    \rho \frac{\partial \mathbf{v}}{\partial t} + \nabla \cdot \left( \rho \mathbf{v} \mathbf{v} + P \right) = \rho \mathbf{g} - \mathbf{G}_\mathrm{r},
\end{equation}
\begin{equation}
    \frac{\partial E}{\partial t} + \nabla \cdot \left[ \left( E + P \right) \mathbf{v} \right] = \rho \mathbf{v} \cdot \mathbf{g} - G_\mathrm{r}^0 + \rho \epsilon_{\mathrm{CR}},
\end{equation}
\begin{equation}
    \frac{\partial I}{\partial t} + c\,\mathbf{n} \cdot \nabla I = S \left( I, \mathbf{n} \right).
\end{equation}
Here, $\mathbf{g}$ is the gravitational acceleration, and $\epsilon_\mathrm{CR}$ is the cosmic-ray heating rate per unit mass of gas. $\mathbf{G_\mathrm{r}}$ and $G_\mathrm{r}^0$ are respectively the radiative force and heating/cooling term that couple the hydrodynamics with the radiative transfer equation. The total energy density $E$ is related to the gas pressure $P$ by
\begin{equation}
    E = \frac{P}{\gamma - 1} + \frac{1}{2} \rho \mathbf{v}^2,
\end{equation}
where $\gamma = 5/3$ is the adiabatic index. The source term $S \left( I, \mathbf{n} \right)$ in the radiative transfer equation (9) is defined by
\begin{equation}
    S \left( I, \mathbf{n} \right) = c \rho \left[ \kappa_\mathrm{a,P} \left( \frac{c a_\mathrm{r} T^4}{4 \pi} - J \right) + \kappa_\mathrm{a,R} \left( J - I \right) \right],
\end{equation}
where $\kappa_\mathrm{a,P}$ and $\kappa_\mathrm{a,R}$ are the Planck and Rosseland mean opacity, $T = \rho k_\mathrm{B} T / \mu$ is the gas temperature calculated with the mean molecular weight $\mu = 2.33 \mathrm{\,u}$, and $J = \int I d\Omega / 4\pi$ is the mean intensity. The rest of the symbols carry their usual meanings.

The set of equations is solved in spherical coordinates assuming 1-D symmetry in the \textit{Athena++} framework \citep{stone20}. The radiative transfer equation is solved as described by \citet{jiang19} except that the equation is solved covariantly as described by \citet{chang20} with additional geometric terms to account for spherical geometry following \citet{davisandgammie20}.

The RHD simulation starts with a stationary cloud core with uniform density and temperature. A setup similar to model M1a from \citet{masunaga98} is adopted. The computational domain from $r = 0.1$ to $10^4 \mathrm{\,AU}$ is resolved by 128 logarithmically spaced cells. A total mass of $1 \mathrm{\,M_\odot}$ with an initial temperature of $T_0 = 8 \mathrm{\,K}$ \citep[as opposed to 10K in][]{masunaga98} is enclosed in the domain. To maintain constant temperature in the initial isothermal phase, the cosmic-ray heating rate $\epsilon_\mathrm{CR}$ is set to $c \kappa_{a,P} a_r T_0^4$ where the opacity $\kappa_{a,P}$ is evaluated at the initial density $\rho_0$ and temperature $T_0$ to account for the radiative cooling. We adopted the same density- and temperature-dependent prescription for frequency-averaged opacity as in \citet{kuiper10}. To speed up the simulation without sacrificing accuracy, the reduced speed of light approximation~\citep[see, e.g., ][]{chang20} with a reduction factor of $10^4$ is used. Reflecting boundary conditions are employed at both the inner and outer radial boundaries for hydrodynamic variables to prevent gas from exiting the computational domain and conserve the total mass. This is a good approximation at the inner boundary because we are interested in the early phase of the star formation up to the thermally supported FHSC, before the formation of a central protostellar object. Reflecting and vacuum boundary conditions are implemented at the inner and outer boundaries respectively for radiation to both prevent radiation from leaking through the inner boundary while allowing radiation to exit through the outer boundary. The system is evolved for 255.1 kyr, up to the moment when the central temperature reaches $2000 \mathrm{\,K}$, at which point the dissociation of molecular hydrogen that triggers the collapse of the FHSC is expected. However, it should be noted that the chemical modeling is performed for infalling parcels for which final temperatures do not exceed 400 K, beyond which the description of the chemistry could become inaccurate, due in part to some of the measured gas-phase reaction rate coefficients used in the network exceeding their recommended temperature range. Thus, 1 AU is the minimum radius resolved within the chemo-dynamical model at the end of the dynamical evolution (which corresponds to a parcel of gas with a minimum radius of 2700 AU when the collapse begins). This is small enough to resolve the effective radius of the FHSC ($\thicksim$5 AU). Overall, the dynamical results are consistent with \citet{masunaga98}.

\section{Results \& Discussions}

\subsection{Physical/dynamical behavior}
\begin{figure*}
\gridline{\fig{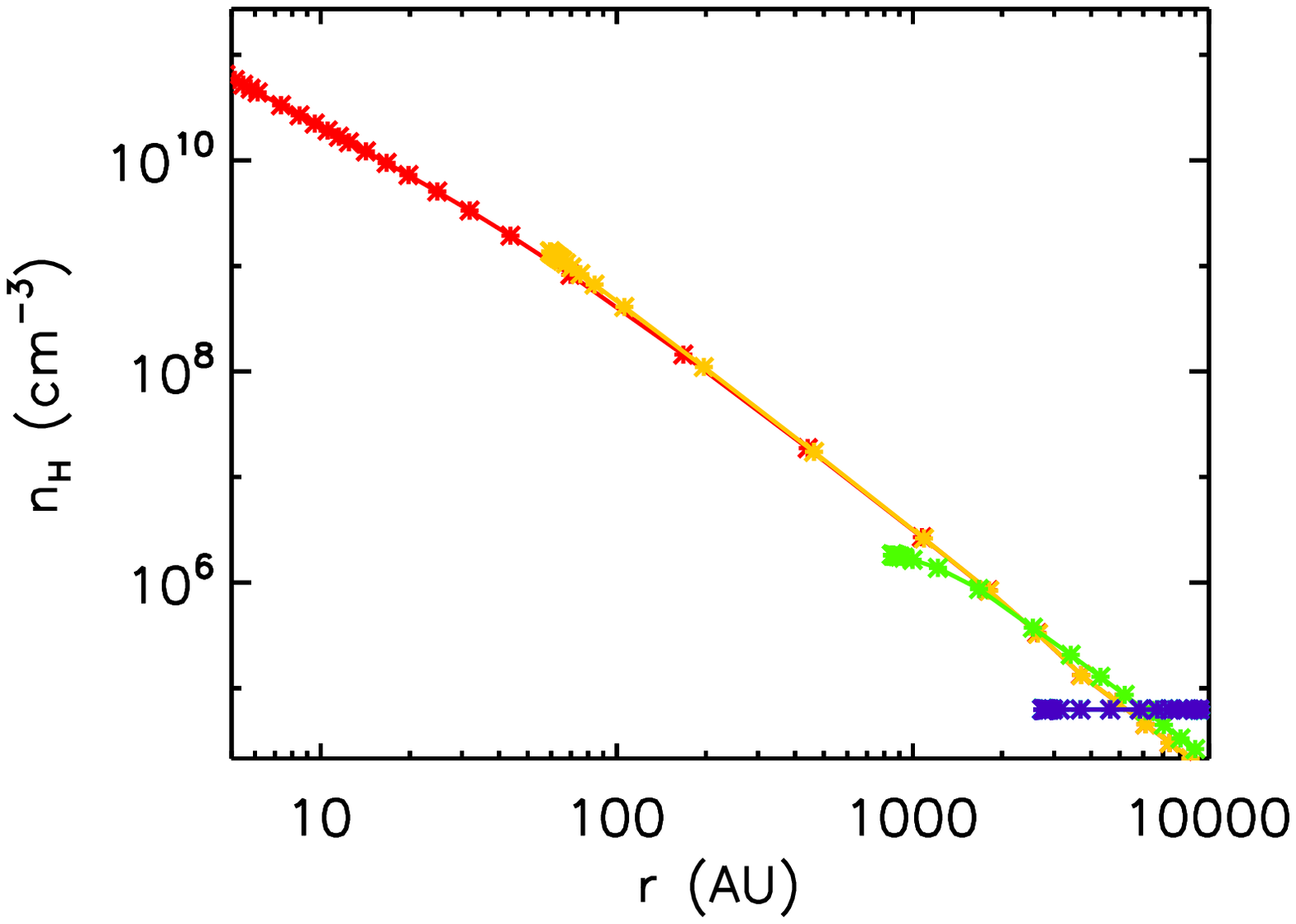}{0.5\textwidth}{}
          \fig{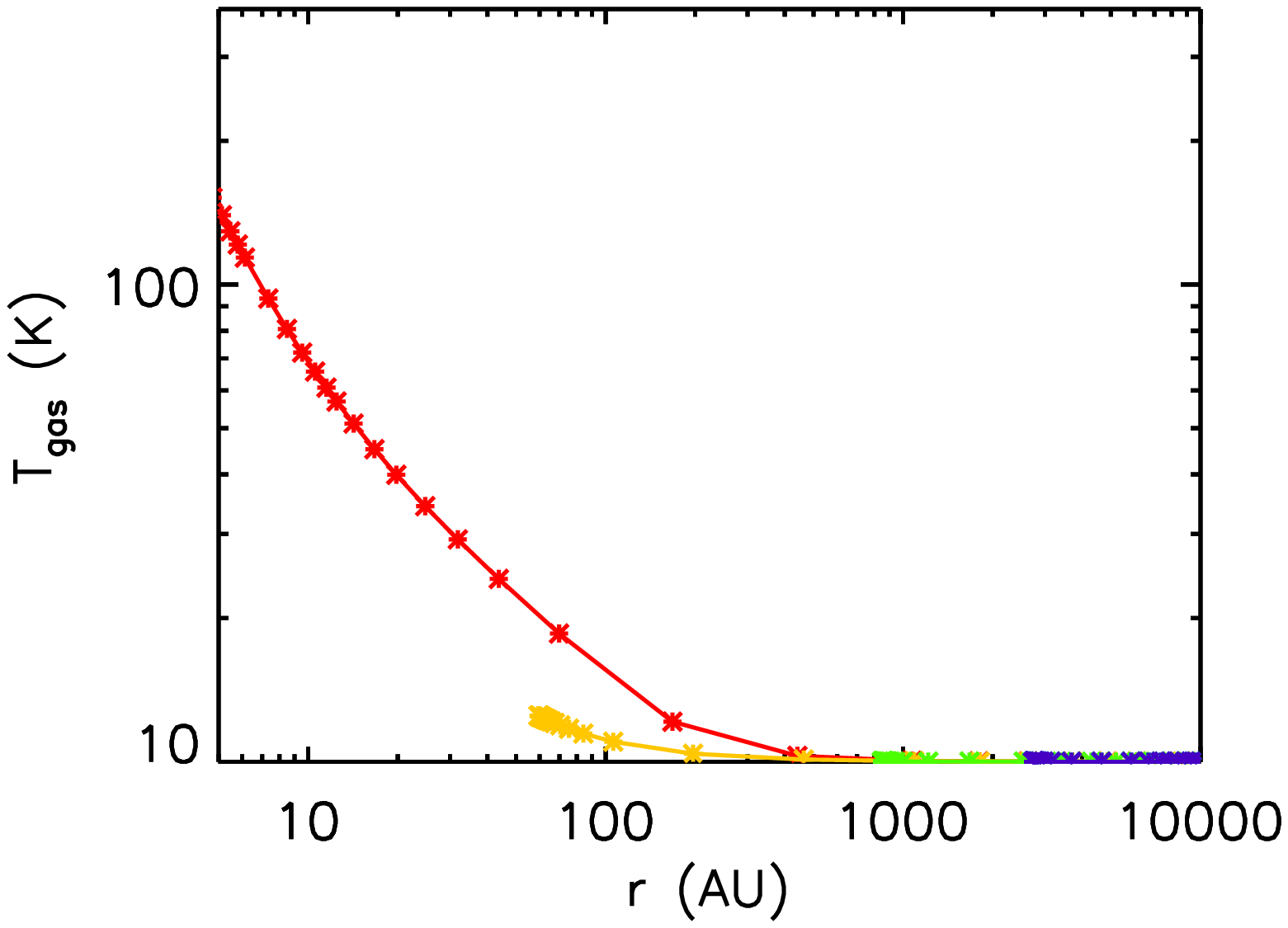}{0.5\textwidth}{}}
   \caption{Snapshots of the physical conditions as functions of radius at four different evolutionary times in the RHD model ; 1 yr (blue), 192,400 yr (green), 254,800 yr (yellow), and 255,100 yr (red)
   \label{fig:physcon_snapshots}}
\end{figure*}

\begin{figure}[h] 
   \centering
   \includegraphics[width=4in]{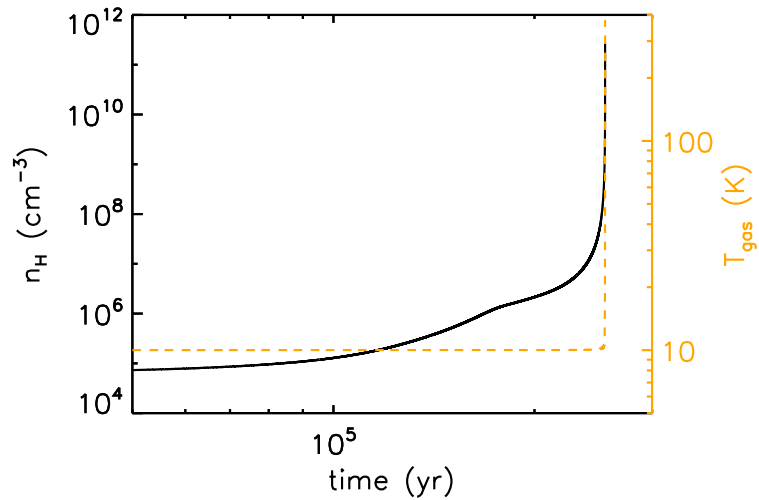} 
   \caption{Physical evolution of the innermost trajectory as a function of time during the RHD model. The black solid line and the yellow dashed line indicates the evolution of density and temperature, respectively.
   \label{fig:innermost_phys_evol}}
\end{figure}

Figure~\ref{fig:physcon_snapshots} shows snapshots of the gas density and temperature as functions of radius at four different evolutionary times in the RHD model of FHSC formation; these correspond to a moment early in the simulation, to the end-time of the simulation, and to two intermediate moments during the collapse, which becomes rapid at late times. The collapsing cloud core is initially optically thin to the thermal emission from dust grains, and it collapses isothermally for a long evolutionary time because the compressional heating rate is much smaller than the radiative cooling rate. During this time, the density develops an $r^{-2}$ profile, which resembles the Larson-Penston solution for the isothermal infalling envelope~\citep{larson69, penston69}. As the density evolves, the collapsing cloud becomes optically thick, forming the adiabatic FHSC in the central region. Once the adiabatic core is formed, the warm-up timescale is very short, showing a step-like profile for the temperature as a function of time (see Figure~\ref{fig:innermost_phys_evol}, which indicates the temperature behavior for an individual trajectory). Considering that the second collapse (which terminates the FHSC phase) begins when the central temperature reaches about 2000 K, such rapid evolution of temperature explains well the short lifetime of FHSC. This process is discussed in further detail by \citet{masunaga98}.

\subsection{Main ice constituents} \label{subsec:discussion-main_ice}

\begin{figure*}
\gridline{\fig{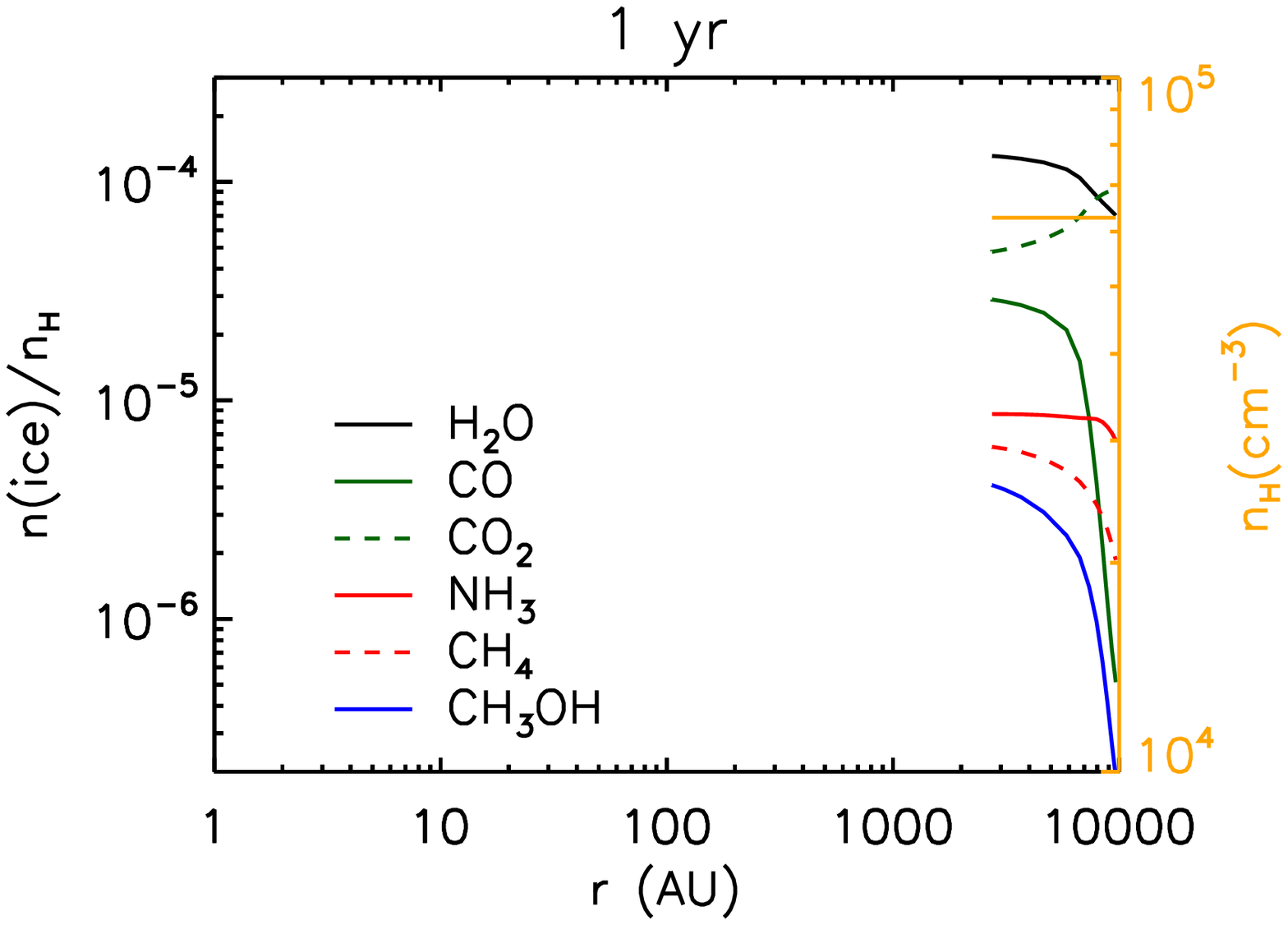}{0.5\textwidth}{}
          \fig{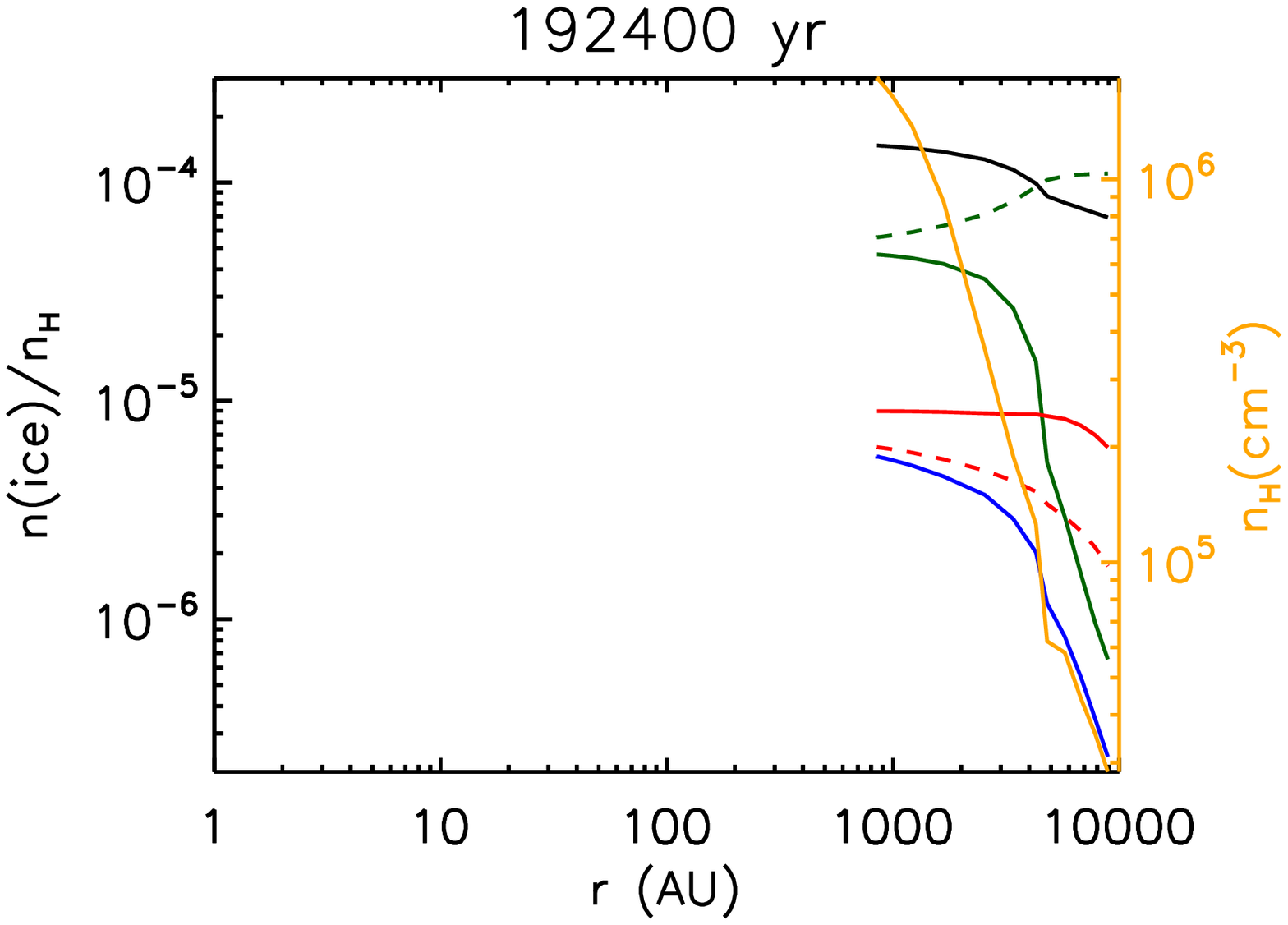}{0.5\textwidth}{}}
\gridline{\fig{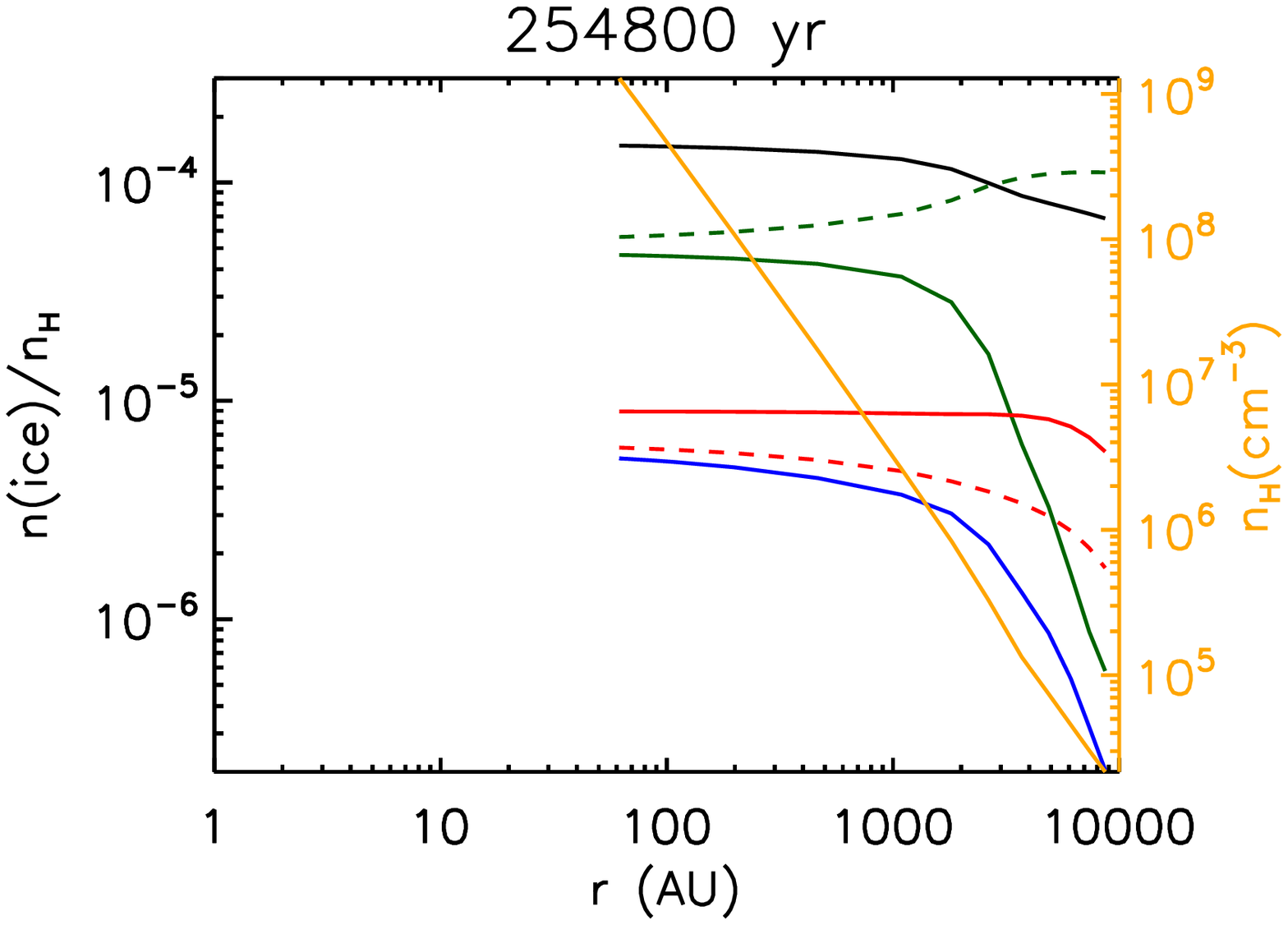}{0.5\textwidth}{}
          \fig{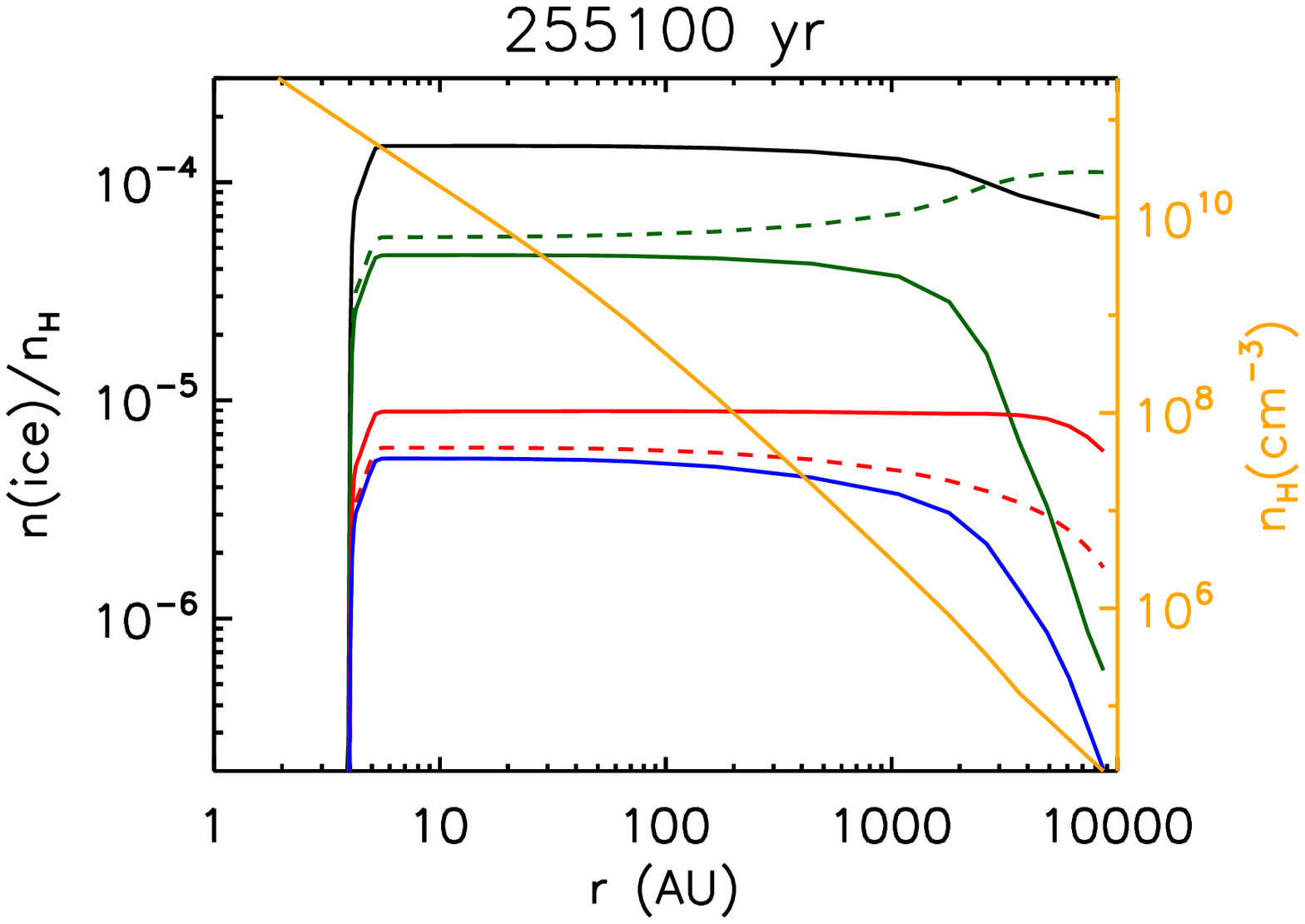}{0.5\textwidth}{}}
   \caption{Radial distribution of the main ice constituent abundances during the evolution of Cha-MMS1; $\ce{H2O}$ (solid black), $\ce{CO}$ (solid green), $\ce{CO2}$ (dashed green), $\ce{NH3}$ (solid red), $\ce{CH4}$ (dashed red) and $\ce{CH3OH}$ (solid blue). The yellow lines indicate density profile within the source obtained from the RHD simulation.
   \label{fig:main_rVSXi_snapshots}}
\end{figure*}

Figure~\ref{fig:main_rVSXi_snapshots} shows the four snapshots of the radial distribution of main ice constituents at the same evolutionary times presented in figure~\ref{fig:physcon_snapshots}, i.e. 1, 192,400, 254,800, and 255,100 yr. Abundances are shown as a fraction of total hydrogen in the gas. Although ice species are formed throughout the whole collapsing core, the chemical evolution is presented only for the infalling parcels with initial radius $r_\textrm{init}~>$ 2700 AU; this ensures that their final radii are no smaller than $\sim$1~AU, and thus that their final temperatures do not exceed the nominal 400~K maximum temperature of the chemical model.

\begin{figure}[h] 
   \centering
   \includegraphics[width=4in]{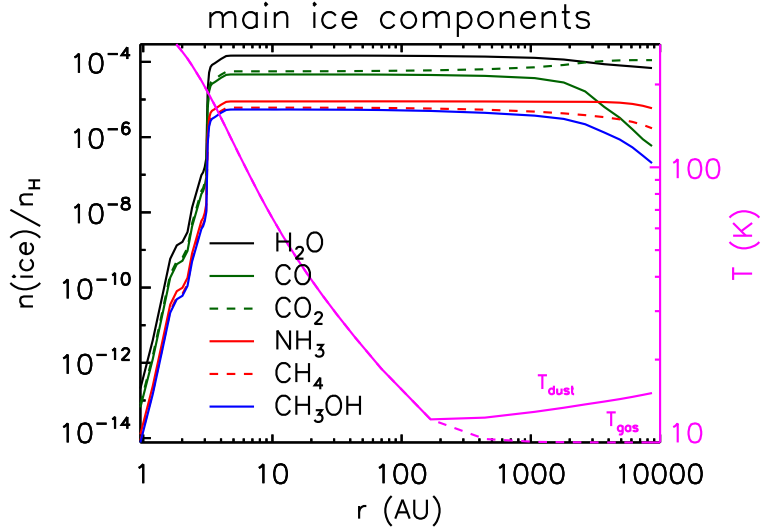} 
   \caption{Final distribution of the main ice constituent abundances within Cha-MMS1; $\ce{H2O}$ (solid black), $\ce{CO}$ (solid green), $\ce{CO2}$ (dashed green), $\ce{NH3}$ (solid red), $\ce{CH4}$ (dashed red) and $\ce{CH3OH}$ (solid blue). The magenta lines indicate the temperature profile within the source obtained from the RHD simulation.
   \label{fig:main_rVSXi}}
\end{figure}

Figure \ref{fig:main_rVSXi} again shows the final spatial distribution of the main ice components, here with the gas and dust temperature profiles shown. 
The dust temperature in the model is consistent with the estimate toward Cha-MMS1 provided by \citet{busch20} ($\thicksim$30K at 55 AU) based on ALMA dust continuum observations.
The divergence between the gas and dust temperatures at radii $>$200 AU occurs because $T_\textrm{gas}$ is determined directly by the RHD simulation, in which only protostellar heating is considered, in combination with a fixed minimum temperature of 10~K; meanwhile $T_\textrm{dust}$ is governed by external heating, as determined by the visual extinction. The two temperatures converge when the RHD-derived temperature exceeds, and thus overrides, the extinction-derived dust temperature, at which point it is assumed that the gas density is already high enough to ensure that the two would be well coupled.
In fact, in the model, this convergence occurs at a rather high density, around $10^{8}$ cm$^{-3}$, while an explicit treatment of the gas-dust thermal coupling would be expected to achieve convergence at a substantially lower value \citep[$\gtrsim 10^{4.5}$ cm$^{-3}$;][]{goldsmith01}. However, the difference between $T_\textrm{gas}$ and $T_\textrm{dust}$ is generally small during this period of divergence, and is due to the use of only an internal heat source in the RHD model (which controls $T_\textrm{gas}$). This has a negligible effect on the dynamics at this temperature (i.e. 10--15~K), while the effect on the gas-phase chemistry is also minor (as reactions with strong temperature dependencies are, in general, already negligibly slow). The focus of our model is the grain-surface and ice chemistry, thus this simplification in the gas temperature treatment at early times/low extinctions is considered acceptable.

The ice abundances within the effective radius of the FHSC~\citep[$\thicksim$ 5 AU;][]{masunaga98} significantly decrease due to active thermal desorption of ice species. The distribution of main ice constituents generally shows a locally-peaked structure, peaking between 10 AU and 1000 AU. This can be explained by a centrally condensed density profile combined with high temperature toward the core center. However, this locally-peaked structure is not always seen. For example, the ice abundance of $\ce{CO2}$ increases with radius, and the abundances of $\ce{NH3}$ and $\ce{CH4}$ in the ice mantles change little outside of the 5 AU effective radius of the FHSC. This implies the involvement of chemistry depending on ice species.

From the abundance data, ice column densities are calculated by integrating the number density (cm$^{-3}$) of each ice species through a line of sight toward the core center (technically, 1AU offset from the core center). In this case, the protostar itself would be the background source; the column densities are thus derived by integrating the number density from the core center to the edge of the core.

Table~\ref{table:main_ice_abun_fnl} summarizes ice abundance with respect to water ice column density toward the core center. The relative abundances reproduce well the main ice composition of clearly identified species from past observations towards low-mass YSOs~\citep{boogert15}. The relative abundances of all main ice species are within the observational range. The CO:\ce{CO2} abundance ratio from the model (0.81) shows a good match with the observations based on median values of CO and \ce{CO2} abundances (0.75), although the relative abundance of \ce{CO2} with respect to water exceeds the upper quartile value from the observations. The broad consistency suggests that the Cha-MMS1 model presented here should provide a reliable basis for the chemical investigation. The peak water ice abundances with respect to total hydrogen ($\thicksim10^{-4}$) in the model~(see figure~\ref{fig:main_rVSXi_snapshots}) are higher than required to reproduce the highest abundances seen in the observations, by a factor of around 2. This suggests that column densities of other ice species could be overproduced by a similar factor. For this reason, we further investigate the model results with a focus on the relative abundances with respect to water ice rather than on the ice abundances relative to atomic hydrogen; the possible causes of the higher ice abundances produced by the model are discussed in \S~\ref{sec:discussion-water}. The influence of this effect on synthesized observational spectra is also tested in \S~\ref{sec:discussion-IR_synthesis}. 

\begin{deluxetable}{cccccc}
\tablewidth{0pt}
\tabletypesize{\footnotesize}
\tablecolumns{6}
\tablecaption{The relative abundances of the main ice constituents with respect to water ice column density \label{table:main_ice_abun_fnl}}
\scriptsize
\tablehead{
\colhead{} & \colhead{CO} & \colhead{$\ce{CO2}$} & \colhead{$\ce{CH4}$} & \colhead{$\ce{NH3}$} & \colhead{$\ce{CH3OH}$} \\
\colhead{} & \colhead{(\%)} & \colhead{(\%)} & \colhead{(\%)} & \colhead{(\%)} & \colhead{(\%)} 
}
\startdata
model & 31.5 & 38.9 & 4.1 & 6.1 & 3.7 \\
observed median value (LYSO) \tablenotemark{a} & 21$^{35}_{12} (18)$ & 28$^{37}_{23}$ & 4.5 $^{6}_{3} (3)$ & 6  $^{8}_{4} (4)$ & 6 $^{12}_{5} (5)$ \\
observed range	& ($<$3) - 85 & 12 - 50 & 1 - 11 & 3 - 10 & ($<$1) - 25\\			 
\enddata
\tablecomments{$^{a}$\citet{oberg11} ; For each molecule, the second row gives the median and lower and upper quartile values of the
detections, and in brackets the median including upper limits. The third row gives the full range of abundances.}
\end{deluxetable}

\begin{figure}[h] 
   \centering
   \includegraphics[width=4in]{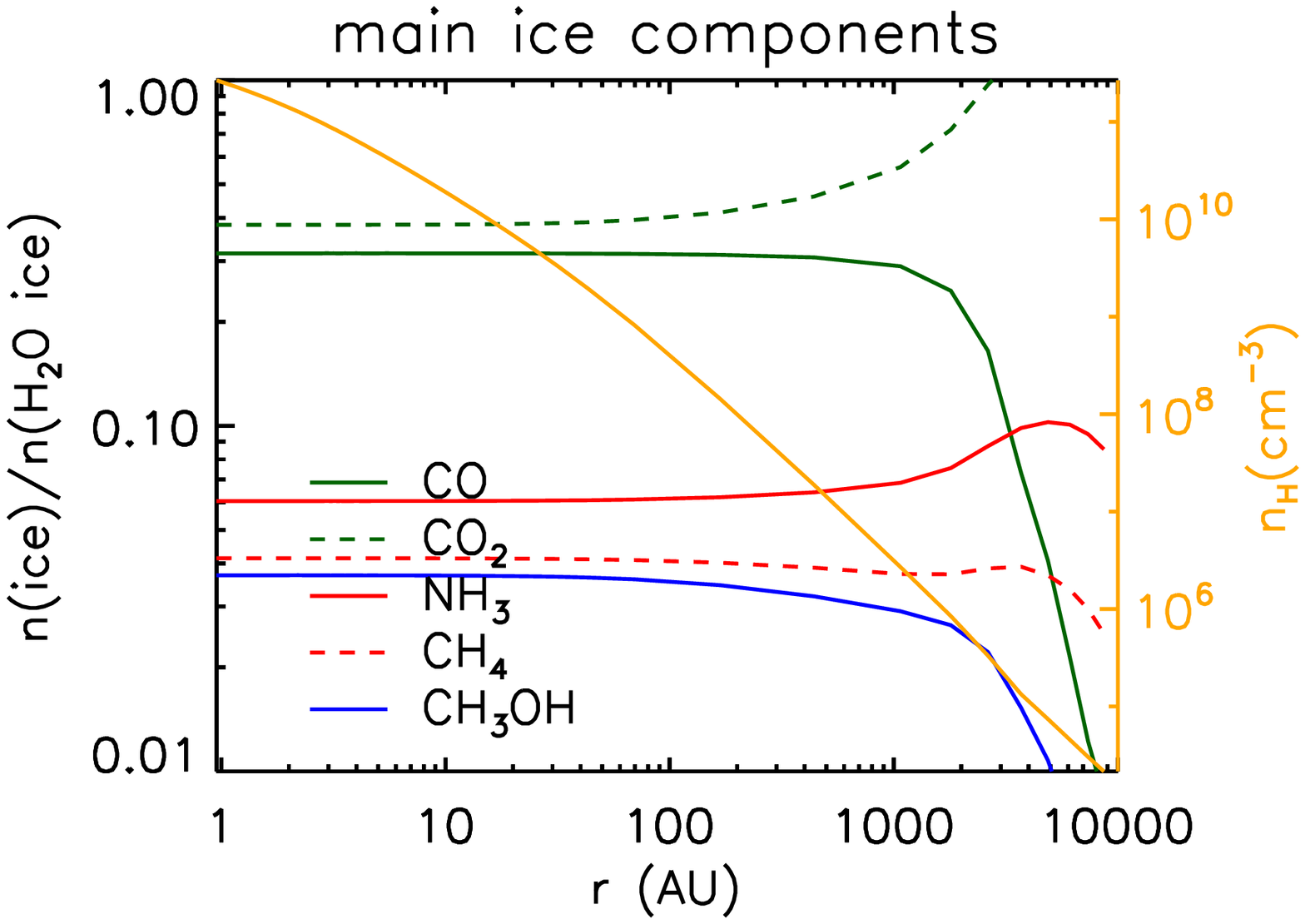} 
   \caption{Final distribution of the main ice constituent abundances w.r.t. water ice within Cha-MMS1; $\ce{H2O}$ (solid black), $\ce{CO}$ (solid green), $\ce{CO2}$ (dashed green), $\ce{NH3}$ (solid red), $\ce{CH4}$ (dashed red) and $\ce{CH3OH}$ (solid blue). The yellow lines indicate the density profile within the source determined by solving the RHD equations.
   \label{fig:main_rVSXiwrtH2O}}
\end{figure}

Figure~\ref{fig:main_rVSXiwrtH2O} shows the radial distribution of abundances for the main ice constituents (with respect to water ice) at the end of the model run, i.e., once the FHSC is formed. The following three features may be identified in the distribution: 

(1) An anti-correlation between CO and \ce{CO2} at r $>$ 2000 AU -- the CO:\ce{CO2} ratio is almost constant ($\thicksim 0.8$) at $r < 100$~AU, while the ratio rapidly decreases in the outer regions. This also results in the \ce{CO2}/\ce{H2O} ratio exceeding 1 at $r \ga 3000$~AU (which is improbable given that no observational sightline has ever revealed such a ratio). The latter effect is related to the efficient conversion of CO to \ce{CO2} (OH + CO $\rightarrow$ \ce{CO2} + H) on grain-surfaces and within ice mantles. On the grain surface, this is a diffusive reaction due to the high dust temperatures at outer regions while nondiffusive reaction dominates within the mantles. Thus, OH radicals produced as a result of PD of water ices quickly react with nearby CO (which is one of the most abundant ice species after water). The rate at which this instantaneous reaction mechanism occurs depends on the photodissociation rate of water ice, thus more efficient conversion appears at outer regions. Although photodissociation of \ce{CO2} acts as a backward reaction to form CO (\ce{CO2} $\rightarrow$ CO + H), the formation of \ce{CO2} through the nondiffusive process dominates over its destruction. Thus, the formation of \ce{CO2} is sensitive to the description of visual extinction in the model, as this parameter governs dust temperature and photodissociation rates at the same time. Our simple assumption for the single isolated core could cause the underestimation of visual extinctions in the outer regions, resulting in the conversion of CO to \ce{CO2} which might be too efficient (see the discussion in \S~\ref{sec:discussion-colden_profile}). However, the effect will have only a very minor influence on the column densities calculated below.

(2) The correlation between CO and \ce{CH3OH} ice distribution; this indicates that \ce{CH3OH} ice is mainly formed by ongoing CO hydrogenation on grain surfaces.

(3) Minor spatial changes in \ce{NH3} and \ce{CH4} ice abundances; the chemistry related to these two saturated species is fairly insensitive to the physical variations with radius. The gradual increases of their ice abundances with respect to water ice seen at outer radii is caused by the faster accretion of \ce{NH3} and \ce{CH4} than \ce{H2O}, as the gas-phase budget of C and N has not run down yet at those outer radii~\citep{garrod21}.

\begin{deluxetable}{cc}
\tablewidth{0pt}
\tabletypesize{\footnotesize}
\tablecolumns{6}
\tablecaption{Ice species investigated in this study
\label{table:ice_spec}}
\scriptsize
\tablehead{
\colhead{category} & \colhead{species}
}
\startdata
main ice constituent & $\ce{H2O}$, $\ce{CO}$, $\ce{CO2}$, $\ce{CH4}$, $\ce{NH3}$, $\ce{CH3OH}$ \\
CO hydrogenation & $\ce{HCO}$, $\ce{H2CO}$, $\ce{CH3O}$, $\ce{CH2OH}$ \\
O-bearing large astronomical species & $\ce{HCOOH}$ (formic acid; FA), $\ce{CH3CHO}$ (acetaldehyde; ACTD), $\ce{CH3OCH3}$ (dimethylether; DME) \\ 
($n_\textrm{atom} \ge 5$) & $\ce{HCOOCH3}$ (methyl formate; MF), $\ce{C2H5OH}$ (ethanol; EtOH), $\ce{CH2OHCHO}$ (glycolaldehyde; GA)\\ 
& $\ce{CH3COOH}$ (acetic acid; AA), $\ce{CH3OCH2OH}$ (methoxy methanol; MM) \\ 
& $\ce{(CH2OH)2}$ (ethylene glycol; EG), \ce{(CH3)2CO} (acetone; ACTN)
\enddata
\end{deluxetable}

\begin{figure*}
\gridline{\fig{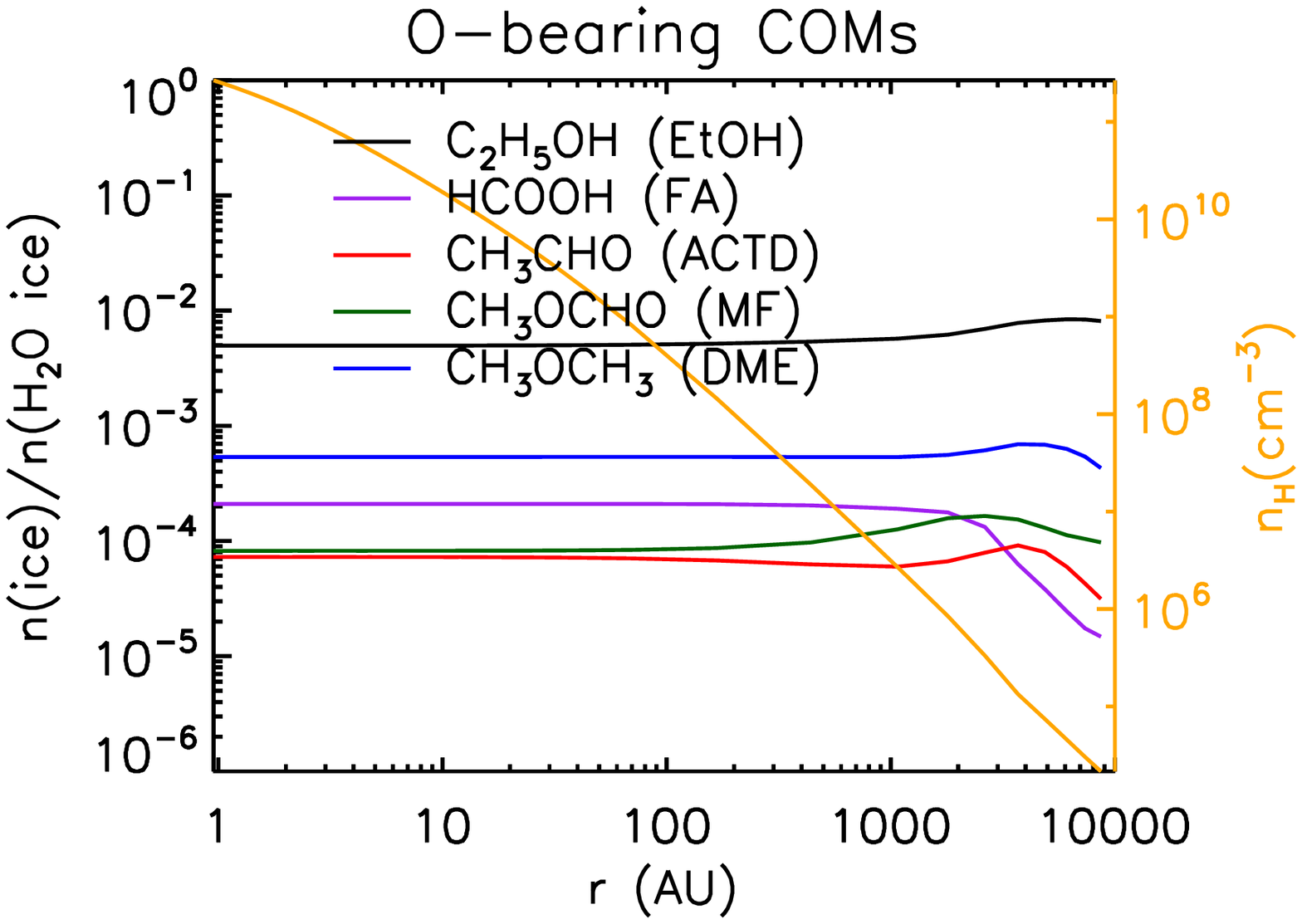}{0.5\textwidth}{}
          \fig{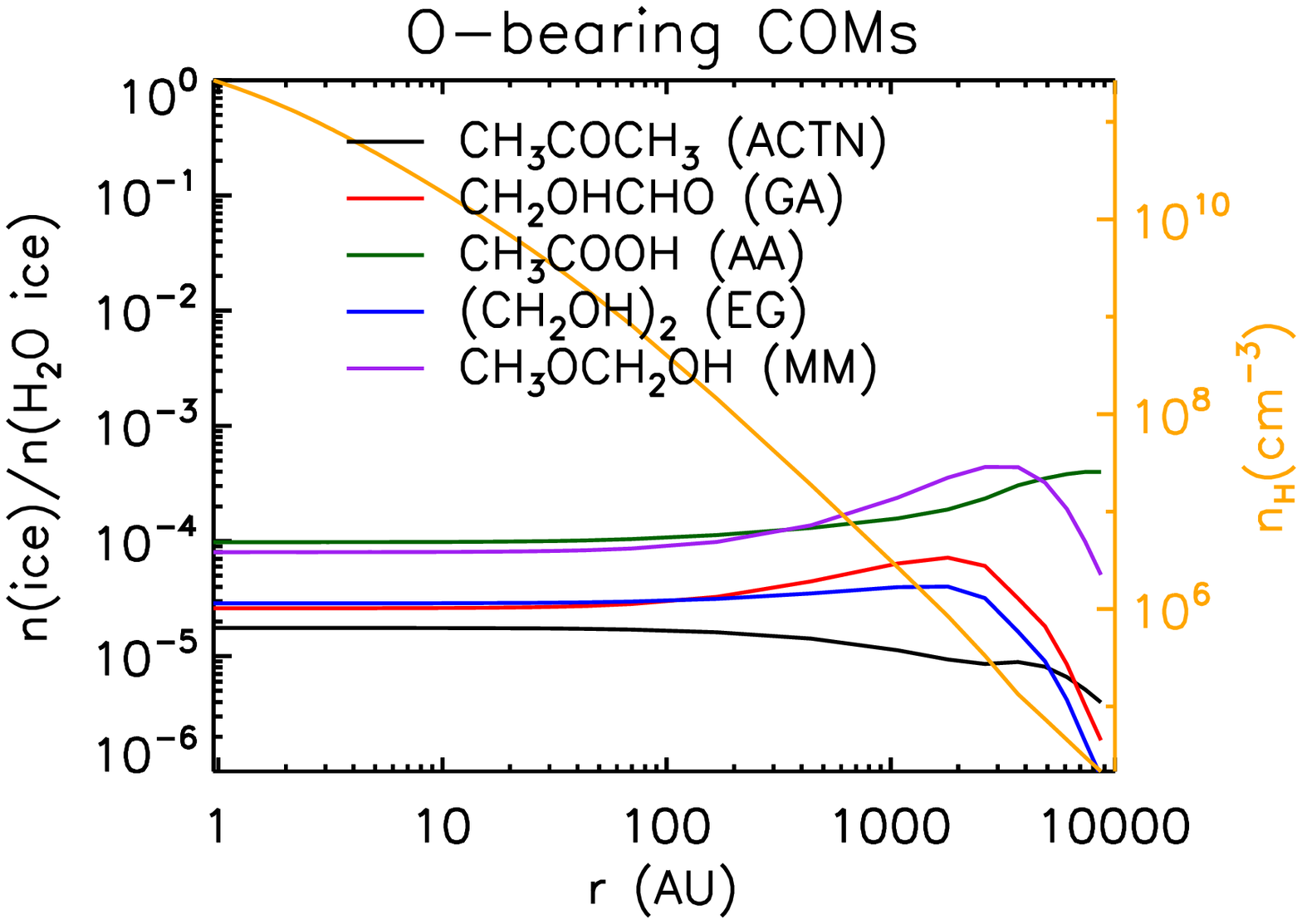}{0.5\textwidth}{}}
\gridline{\fig{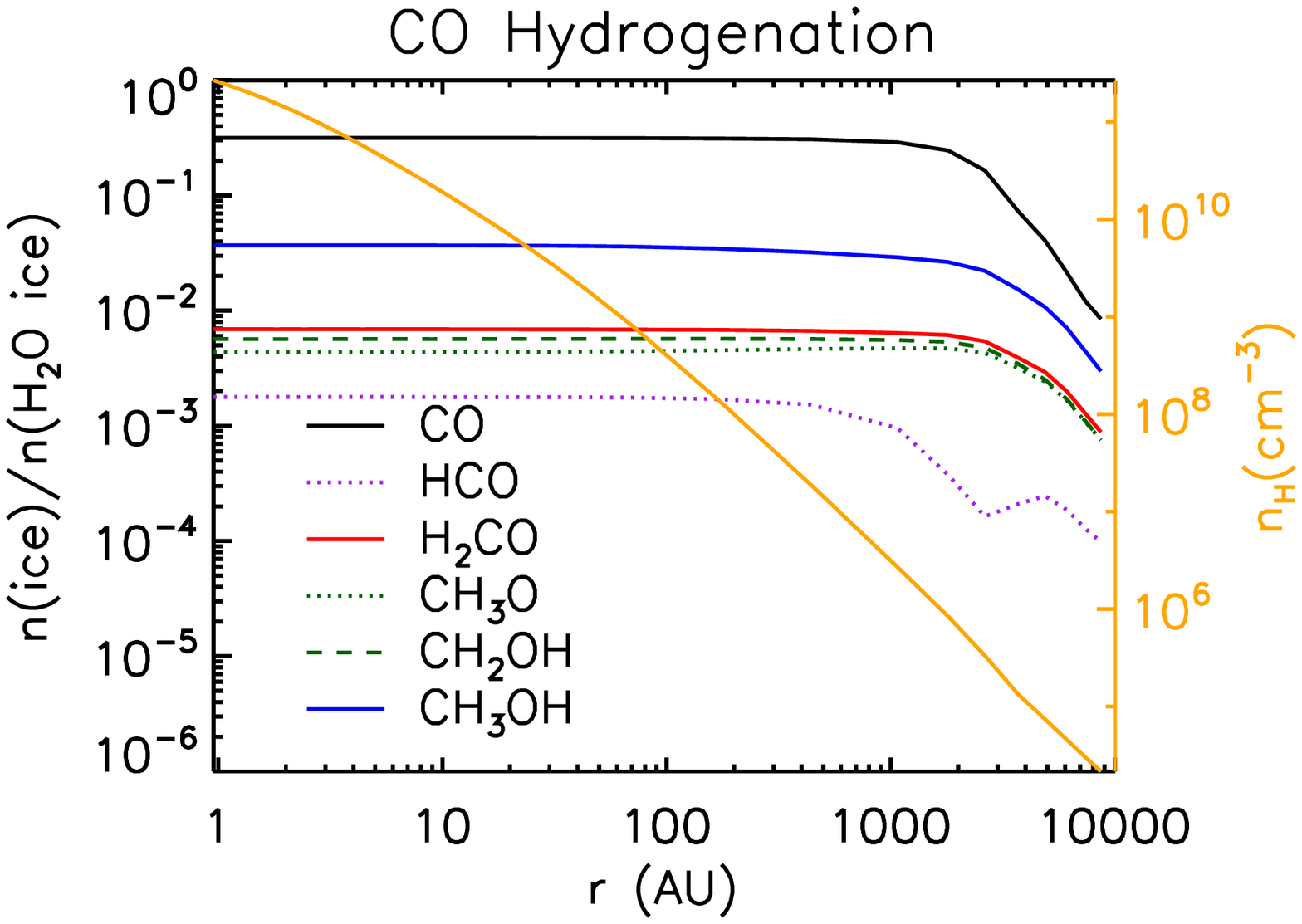}{0.5\textwidth}{}} 
   \caption{Radial distribution of the ice abundances with respect to water ice abundances at the end of the evolution. The yellow lines indicate the density profile within the source determined by solving the RHD equations.
   \label{fig:other_rVSXi}}
\end{figure*}

\subsection{Other volatiles expected to be abundant in the ice mantles} \label{sec:discussion-other_icy_volatiles}

We have investigated other volatiles expected to be abundant in the ice mantles, including CO-hydrogenation intermediates and well known O-bearing hot-core species; species of interest are listed in Table~\ref{table:ice_spec}, along with their abbreviations. Figure~\ref{fig:other_rVSXi} shows the spatial distribution of these ice species with respect to the local fractional abundance of water ice, $X$(H$_2$O), at the end of the evolution. The two upper panels show large ($n_\textrm{atom} \ge 5$) O-bearing species that have been detected toward hot cores while the lower panel shows the intermediates related to CO hydrogenation. 

The results from this model predict high ice abundances for the CO-hydrogenation intermediates, including radicals ($\ga~10^{-3}$ $\times$X(H$_2$O)) although this value is considerably lower than the upper limit of the storage capacity of the ice for reactive radical species~\citep[$< 10\%$;][]{golden58, jacksonandmontroll58, jackson59a, jackson59b}. Garrod et al. (2022) further discuss the abundances of radicals in the ices. For more complex O-bearing species, their solid-phase abundances are typically on the order of $10^{-5}$--$10^{-4}$$\times$X(H$_2$O), while EtOH shows distinctively high solid-phase abundances of $\ga~10^{-3}$$\times$X(H$_2$O). The five O-bearing species presented in the upper-left panel of figure~\ref{fig:other_rVSXi} also show relatively high abundances. Observations of these species have, either directly or indirectly, implied their presence within the ice mantles. Three of them (ACTD, MF, and DME) have been identified in the gas phase in cold, prestellar environments~\citep{bacmann12, vastel14, cernicharo12,scibelli20}, with grain-surface formation being suggested as an explanation. In addition, the analysis of the mid-IR spectrum of the massive protostar W33A -- one of the few sources for which a high quality mid-IR spectrum is currently available -- identified three prominent features that have been attributed to EtOH, FA, and ACTD, although their identification in the ices remains uncertain~\citep{schutte99, boogert08, oberg11}. \citet{terwisscha18} re-analyzed the same spectrum with more accurate spectroscopic laboratory data, 
deriving upper limits for ice abundances with respect to water of 1.9 \% and 2.3 \% for EtOH and ACTD, respectively. This is 1-2 orders of magnitude more than predicted by our model. AA and MM have not been identified from the ice observations so far, but their large abundances in the model suggests them to be plausible candidates for solid-phase detection, their spectral characteristics notwithstanding. In particular, those species could be enhanced at the outer radius (in low-density regions) by almost one order of magnitude compared to the core center.

We categorize the spatial distribution of the species shown in figure~\ref{fig:other_rVSXi} into three types: 1) pudding shape, i.e. flattened center, decreasing at large radii; 2) bowl shape, i.e. flattened center, increasing at larger  radii; and 3) locally-peaked, with $X$(i)/$X$(H$_2$O) peaking at a few thousand AU. Table~\ref{table:cat_spec_spatial_distribution} shows the shape with which we identify each of the ice species shown in the figure. Interestingly, aside from the three aforementioned COMs (ACTD, MF, and DME) detected in cold environments, the chemical species following each type of distribution can be linked back to a specific radical or molecule involving their formation. For example, the CO ice distribution is a pudding shape, and most of the ice species whose formation is directly related either to the CO$\rightarrow$FA conversion pathway (see equations 1 and 2) or to the hydrogenation products of CO show similar distributions.

As for the larger species, whose formation pathways are more complex, the abundance profiles are nevertheless found to depend on which functional groups are present. All species with bowl-shaped distributions contain a methyl group (\ce{-CH3}), while all species containing the hydroxymethyl radical (\ce{-CH2OH}) exhibit locally-peaked distributions peaking at a few thousand AU. ACTD, MF, and DME show locally-peaked distributions despite each containing a methyl group, but in the test model of a moderately higher extinction environment their distributions converge to the bowl-shape distribution. This result shows that \ce{-CH3} (which perhaps may include \ce{-CH3O}) and \ce{-CH2OH} might be the key functional groups determining the spatial profiles of some solid-phase species. A recent high-resolution observation indirectly implies this; gas-phase observation of O-bearing COMs in Orion KL show different morphology depending on which functional group is present in the molecule~\citep{tercero18}. To explain this, \citet{tercero18} propose that different radicals, methoxy (\ce{-CH3O}) and hydroxymethyl (\ce{-CH2OH}), may dominate, driving the different chemical complexity of the regions.

Another main factor that determines the radial distribution of ice species is the $A_\textrm{V}$ distribution, dominantly during the pre-model. At the beginning, the visual extinction ranges from 3.0 mag (innermost) to 2.5 mag (outermost), and it evolves to the ranges from 6.2 mag to 2.7 mag with the density evolution during the pre-model. The chemical differentiation of many species (e.g. EtOH) are determined by the balance between formation and destruction processes that both become more efficient further out from the central core. In outer regions, low $A_\textrm{V}$ and/or higher dust temperatures can drive the formation of these species through PDI nondiffusive reactions and/or diffusive reactions, while destruction via PD also becomes active. However, the formation of some species such as MF becomes rather efficient when the density is higher, indicating that the $A_\textrm{V}$ has increased as well. The formation of this type of species is not strongly dependent on the PDI processes in the bulk ice, but on the build up through direct surface production via 3-B nondiffusive reactions, when radicals become more abundant on the grain surfaces~(see figure 13 in Garrod et al. 2021).

\begin{deluxetable}{cc}
\tablewidth{0pt}
\tabletypesize{\footnotesize}
\tablecolumns{2}
\tablecaption{Spatial distribution of ice abundance of COMs w.r.t. water ice within Cha-MMS1
\label{table:cat_spec_spatial_distribution}}
\scriptsize
\tablehead{
\colhead{shape} & \colhead{species}
}
\startdata
pudding & \ce{CO}, \ce{HCO}, \ce{H2CO}, \ce{CH3O}, \ce{CH2OH}, \ce{CH3OH}, \ce{HCOOH} \\
bowl & \ce{CH3COOH}, \ce{C2H5OH}\\
locally-peaked & \ce{CH2OHCHO}, \ce{CH3OCH2OH}, \ce{(CH2OH)2}, \ce{HCOOCH3}, \ce{CH3OCH3}, \ce{CH3CHO}
\enddata
\end{deluxetable}

\begin{figure*}
\gridline{\fig{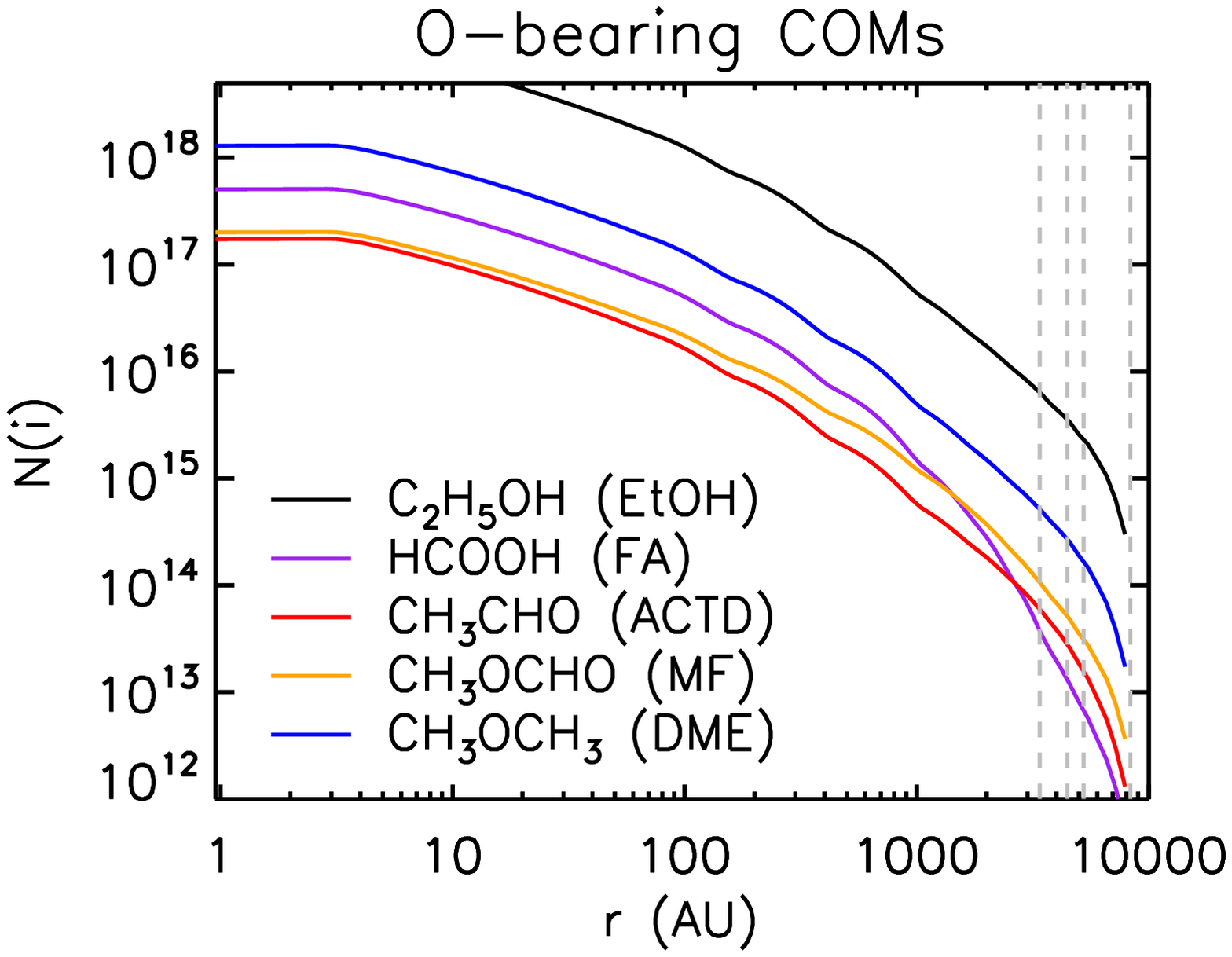}{0.5\textwidth}{}
          \fig{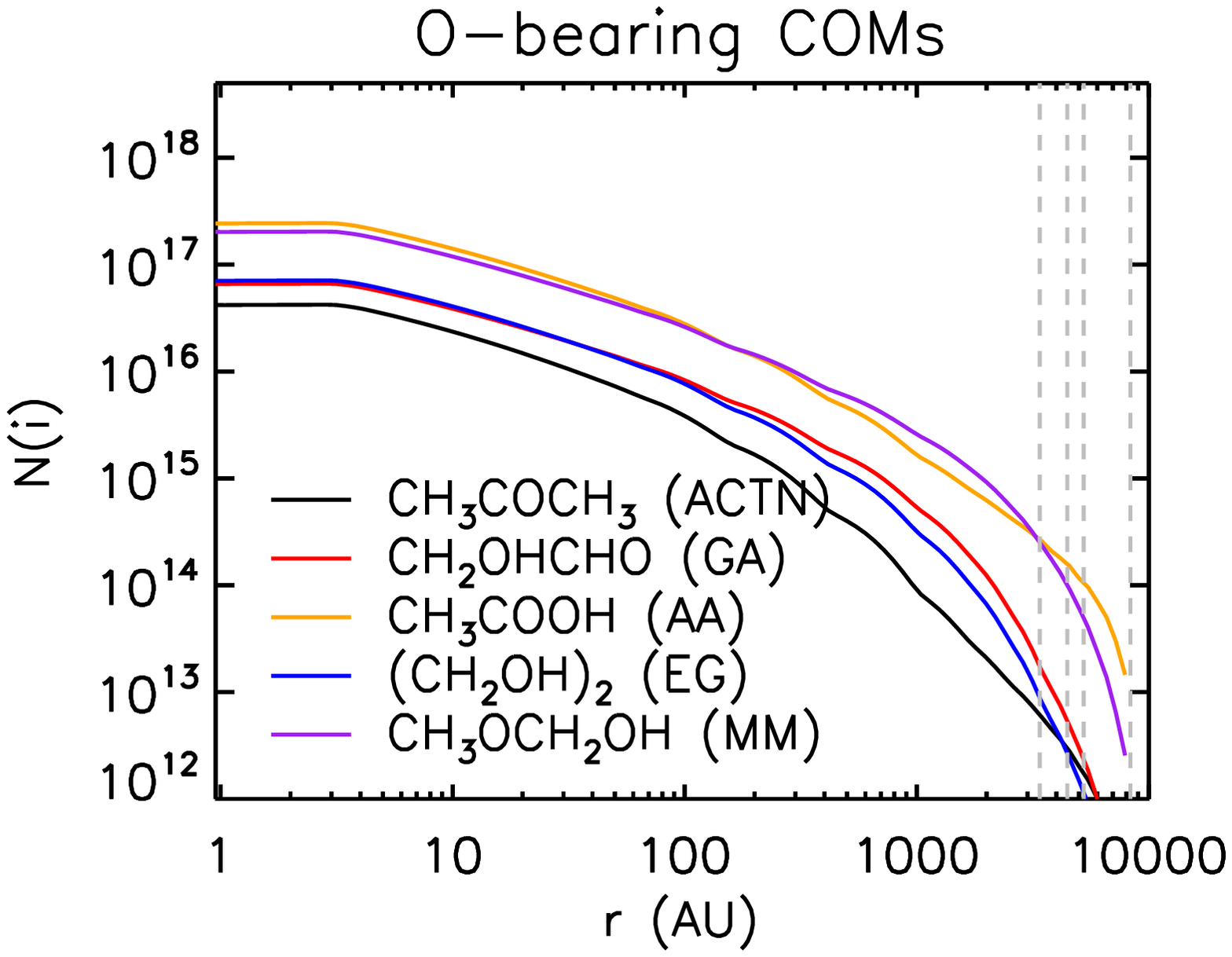}{0.5\textwidth}{}}
   \caption{Radial distribution of the column density produced by our model of Cha-MMS1 for large O-bearing ice species. Gray dashed lines indicate the observed radial offsets from the center of Cha-MMS1 for the lines of sight toward a selection of background stars.
   \label{fig:other_rVSNi}}
\end{figure*}

\begin{figure}[h] 
   \centering
   \includegraphics[width=4in]{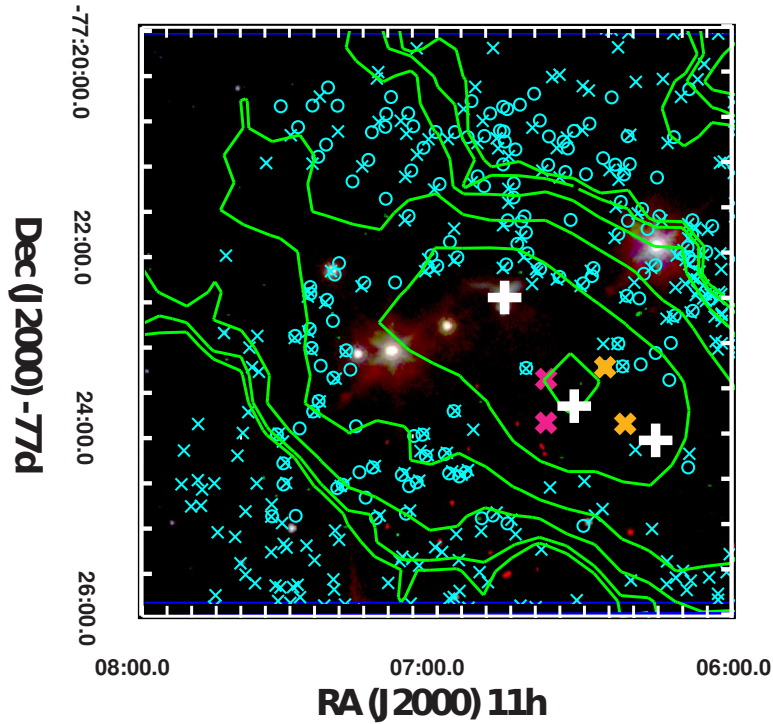} 
   \caption{Map field of ice mapping in the ice age project. Green contours indicates high-$A_\textrm{V}$ 870 $\micron$ dust continuum emission~\citep{belloche11}. The targets of ice age are denoted as large white crosses. From the left, Ced 110 IRS4 (Class I protostar), Cha-MMS1 (Class 0 protostar), and C2 (prestellar core). 
   }
   \label{fig:iceage_mapfield}
\end{figure}

\subsection{Calculated ice column densities} \label{sec:discussion-colden_profile}
Figure~\ref{fig:other_rVSNi} shows the (absolute) column density profiles produced by our model of Cha-MMS1 for large solid-phase molecules. These are calculated by integrating the number density of each species along parallel lines of sight intercepting different offsets from the core center. For comparison, the gray dashed lines indicate the observed radial offsets from the center of Cha-MMS1 of the lines of sight toward a selection of background (BG) stars. The information of these BG stars is listed in table~~\ref{table:bgstar_info}. These lines of sight are showing high-$A_\textrm{V}$ in the {\em Ice Age} field. Considering the high $A_\textrm{V}$ threshold expected for the detection of complex ice species, these lines of sight would be the most probable to identify icy COMs, of those BG stars yet known that lie near on the sky to Cha-MMS1. The two outer BG stars (orange crosses in figure~\ref{fig:iceage_mapfield} -- NIR 38 and J110621.63-772354.1 -- will be observed both with NIRSPEC's fixed slit mode (0.6 to 5 microns) and MIRI's low resolution spectroscopy (LRS) mode (5-14 microns) which operate in the near-IR and mid-IR, respectively. The core center and the other two inner positions (pink crosses in figure~\ref{fig:iceage_mapfield})-- J110637.01-772323.0 and J110637.34-772352.7 -- will be observed with NIRCAM / Wide Field Slitless Spectroscopy (WFSS) only (2.4 to 5 \micron). The combination of these five lines of sight will provide a spatial profile of ice composition possibly including COMs.

The extinction towards the two MIRI LRS targets, NIR 38 and SSTSL2J110621.63-772354.1, was calculated for the {\em Ice Age} ERS program using the observed photometric color taken between the K bandpass at 2.2 $\micron$ and Spitzer IRAC band 1 at 3.6 $\micron$ taken from \citet{persi01} and the IPAC Gator database. The calculation assumed an intrinsic K-IR1 color for background giant stars of 0.03 from \citet{majewski11} and the highest extinction $A_{\lambda}$/$A_\textrm{V}$ curve for molecular clouds from \citet{mcclure09}. The other two background stars were only observed at WISE bands 1 and 2. Therefore we assumed an intrinsic background giant W1-W2 color of -0.08 from \citet{schlafly16} and the $A_\textrm{W1}$/$A_\textrm{W2}$ extinction of 0.463 from \citet{xue16}. From comparing the variation in $A_\textrm{V}$ derived for other, less reddened stars from both this set of colors and other colors, we estimate a conservative 20\% systematic uncertainty in our final $A_\textrm{V}$ measurements.

To see how well the density structure described in the model matches with the observations, an observational column density of total hydrogen toward each background star is compared with a calculation of the column density along a line of sight in the model that lies at the appropriate radial separation from the core center. A similar comparison is made between the observed column density of total H toward the center of Cha-MMS1 and the same line of sight toward the model of the source.

For each line of sight, the column density of total hydrogen, $N_{\textrm{H,obs}}$, is calculated based on the observational visual extinction information, using the relation $N_{\textrm{H,obs}}$=$1.8\times10^{21}A_\textrm{V}~\textrm{cm}^{-2}$. To take some account of observational effects in the comparison, the corresponding model values, $N_{\textrm{H,model}}$, are obtained by convolving the column density profiles with the spatial resolution of Spitzer IRAC (2$\arcsec$), using a Gaussian beam profile.

For the core center, $N_{\textrm{H,obs}}$ derived from 870 $\micron$ dust continuum is also available from the literature~\citep{belloche11}. This value is compared with the model by convolving the modeled column density profile with the beam size of the dust continuum map ($\theta_\textrm{HPBW}$=21.2 $\arcsec$). As seen in table~\ref{table:bgstar_info}, the column density derived from the model is roughly consistent with the observation at the core center. This result remains consistent even when an alternative estimate of the $A_\textrm{V}/N_\textrm{H}$ factor is employed. Based on the study by \citet{chapman09}, around two times lower conversion factor ($N_{\textrm{H,obs}}$=$1.1\times10^{21}A_\textrm{V}~\textrm{cm}^{-2}$) might be appropriate for the high density molecular gas. The employment of this alternative conversion factor results in around two times lower $N_\textrm{H,obs}$, which in fact agrees even better with the model at the core center. However, the disparity in total column density between model and observation grows at larger radii, where $N_{\textrm{H,model}}$ falls more than two orders of magnitude below the observational value for the outermost BG position.

The main reason for this discrepancy is that the physical evolution of Cha-MMS1 in the model is described as a single isolated object with the RHD simulation, while Cha-MMS1 is actually one of nine young, low-mass star-forming cores within a distance of 0.2 pc in the plane of the sky~\citep{maureira20}. For example, another {\em Ice Age} target -- prestellar core C2 (right white cross in figure~\ref{fig:iceage_mapfield})-- exists near the outermost BG position, and it is even closer to this BG position than Cha-MMS1. This explains the much higher column density toward this line of sight in the observation than the model. Thus, the model is unlikely to produce meaningful predictions for ice abundances for the nearest BG stars yet detected in the vicinity of Cha-MMS1. However, it is expected that the JWST observations will reveal new background stars that are closer to the core center and therefore may provide a better constraint on the model.

\begin{deluxetable}{cccccccc}
\tablewidth{0pt}
\tabletypesize{\footnotesize}
\tablecolumns{8}
\tablecaption{The information on background stars near to Cha-MMS1~\label{table:bgstar_info}}
\scriptsize
\tablehead{
\colhead{source name} & \colhead{RA} & \colhead{Dec} & \colhead{radius} & \colhead{$A_\textrm{V}$} & \colhead{$N_{\textrm{H,obs}}$} & \colhead{$N_{\textrm{H,model}}$} & \colhead{$N_{\textrm{H,obs}}/N_{\textrm{H,model}}$}\\
\colhead{} & \colhead{(J2000)} & \colhead{(J2000)} & \colhead{(AU)} & \colhead{(mag)} &\colhead{($10^{23}~\textrm{cm}^{-2}$)} & \colhead{($10^{23}~\textrm{cm}^{-2}$)} & \colhead{} 
}
\startdata
Cha-MMS1 & 11:06:33.46 & -77:23:34.52 & 0 & $>$100\tablenotemark{a} & $>$1.80 & 5.2 &  $>$0.3 \\
 & & & & & 0.92 (870$\micron$ 21.2$\arcsec$)\tablenotemark{b} & 0.26 & 4 \\
J110637.01-772323.0 \tablenotemark{c} & 11:06:37.74 & -77:23:23.83 & 3386 & 106 & 1.91 & 0.04 & 48 \\
J110637.34-772352.7 \tablenotemark{c}& 11:06:37.80&  -77:23:52.80 & 4447 & 95 & 1.71 & 0.02 & 86 \\
NIR 38 & 11:06:26.69 & -77:23:18.6 & 5240 & 60 & 1.08 & 0.01 & 108\\
J110621.64-772354.1 & 11:06:21.64 & -77:23:54.12 & 8330 & 95 & 1.71 & $<$ 0.01 & $>$ 171
\enddata
\tablenotetext{a}{M.K. McClure, Priv. Comm.; This estimation is based on the $A_\textrm{V}$ towards the surrounding background stars. There is a systematic uncertainty of $\thicksim$ 20\% on this estiamtion. See \S~\ref{sec:discussion-colden_profile} for more details.}
\tablenotetext{b}{\citet{belloche11} -- $N_{\textrm{H,obs}}$ is derived from the 870 \micron~ dust continuum emission map.}
\tablenotetext{c}{only two WISE photometric bands are available for the estimation of $A_\textrm{V}$.}
\end{deluxetable}

\begin{deluxetable}{cccccc}
\tablewidth{0pt}
\tabletypesize{\footnotesize}
\tablecolumns{6}
\tablecaption{Ice spectral parameters assumed for the synthesis of ice spectrum ~\label{table:ice_features}}
\scriptsize
\tablehead{
\colhead{species} & \colhead{wavenumber} & \colhead{wavelength} & \multicolumn{2}{c}{FWHM} & \colhead{$A\arcmin$} \\
\colhead{} & \colhead{(cm$^{-1}$)} & \colhead{($\micron$)} & \colhead{(cm$^{-1}$)} & \colhead{($\micron$)} & \colhead{(cm molecule$^{-1}$)} 
}
\startdata
\ce{H2O}   &  3280.0 &    3.05 &    335.0 &    0.31 &  2.0E-16\\
           &  1660.0 &    6.02 &    160.0 &    0.58 &  1.0E-17\\     
           &   760.0 &   13.16 &    240.0 &    4.26 &  3.0E-17\\     
CO         &  2140.0 &    4.67 &      5.0 &    0.01 &  1.1E-17\\ 
\ce{^{13}CO}  &  2092.0 &    4.78 &      1.5 &   0.003 &  1.3E-17\\
\ce{CO2}   &  2343.0 &    4.27 &     18.0 &    0.03 &  7.6E-17\\   
           &   660.0 &   15.15 &     18.0 &    0.41 &  1.2E-17\\ 
\ce{^{13}CO2} &  2282.0 &    4.38 &      3.0 &    0.01 &  7.8E-17\\           
\ce{CH4}   &  3010.0 &    3.32 &      7.0 &    0.01 &  7.0E-18\\
           &  1302.0 &    7.68 &      8.0 &    0.05 &  7.0E-18\\ 
\ce{CH3OH} &  3250.0 &    3.08 &    235.0 &    0.22 &  1.1E-16\\ 
           &  2982.0 &    3.35 &    100.0 &    0.11 &  2.1E-17\\ 
           &  2828.0 &    3.54 &     30.0 &    0.04 &  8.0E-18\\ 
           &  1460.0 &    6.85 &     90.0 &    0.42 &  1.0E-17\\ 
           &  1130.0 &    8.85 &     34.0 &    0.27 &  1.4E-18\\ 
           &  1030.0 &    9.71 &     29.0 &    0.27 &  1.4E-17\\    
           &   700.0 &   14.29 &    200.0 &    4.17 &  1.6E-17\\   
\ce{NH3}   &  3375.0 &    2.96 &     45.0 &    0.04 &  2.3E-17\\   
           &  1630.0 &    6.13 &     60.0 &    0.23 &  5.0E-18\\   
           &  1070.0 &    9.35 &     70.0 &    0.61 &  1.7E-17\\   
EtOH       &   707.0 &   14.14 &    157.0 &    3.18 &  9.0E-18\\
	       &   889.0 &   11.25 &     26.0 &    0.33 &  3.8E-18\\
	       &  1034.0 &    9.67 &     34.0 &    0.32 &  1.4E-17\\
	       &  1092.0 &    9.16 &     24.0 &    0.20 &  8.4E-18\\
	       &  1400.0 &    7.14 &    200.0 &    1.03 &  1.8E-17\\
	       &  2805.0 &    3.57 &    145.0 &    0.18 &  3.2E-17\\
	       &  2979.0 &    3.36 &     29.0 &    0.03 &  1.4E-17\\
	       &  3314.0 &    3.02 &    306.0 &    0.28 &  1.3E-16\\
ACTD       &  3416.0 &    2.93 &     40.0 &    0.03 &  6.3E-19\\
	       &  2917.0 &    3.43 &     17.0 &    0.02 &  3.8E-19\\
	       &  2858.0 &    3.50 &     53.0 &    0.06 &  5.4E-18\\
	       &  1768.0 &    5.66 &     32.0 &    0.10 &  1.3E-18\\
	       &  1721.0 &    5.81 &     27.0 &    0.09 &  3.0E-17\\
	       &  1428.0 &    7.00 &     50.0 &    0.25 &  1.1E-17\\
	       &  1350.0 &    7.41 &     25.0 &    0.14 &  7.1E-18\\
	       &  1122.0 &    8.91 &     30.0 &    0.24 &  5.3E-18\\
ACTN       &  2950.0 &    3.39 &    150.0 &    0.17 &  3.9E-18\\
	       &  1710.0 &    5.85 &     30.0 &    0.10 &  2.7E-17\\
	       &  1440.0 &    6.94 &     60.0 &    0.29 &  9.2E-18\\
	       &  1358.0 &    7.36 &     22.5 &    0.12 &  1.4E-17\\
	       &  1230.0 &    8.13 &     20.0 &    0.13 &  7.4E-18\\
	       &  1090.0 &    9.17 &     10.0 &    0.08 &  1.6E-18\\
	       &   890.0 &   11.24 &     50.0 &    0.63 &  8.3E-19\\
	       &   790.0 &   12.66 &     30.0 &    0.48 &  1.6E-19\\
	       &   535.0 &   18.69 &     10.0 &    0.35 &  2.1E-18\\
DME        &  2985.0 &    3.35 &     93.0 &    0.10 &  2.2E-17\\
	       &  2814.0 &    3.55 &     25.0 &    0.03 &  6.8E-18\\
	       &  2075.0 &    4.82 &     40.0 &    0.09 &  3.7E-19\\
	       &  2002.0 &    5.00 &     28.0 &    0.07 &  3.0E-19\\
	       &  1459.0 &    6.85 &     50.0 &    0.23 &  4.8E-18\\
	       &  1248.0 &    8.01 &     23.0 &    0.15 &  7.8E-19\\
	       &  1163.0 &    8.60 &     23.0 &    0.17 &  1.1E-17\\
	       &  1093.0 &    9.15 &     20.0 &    0.17 &  9.0E-18\\
	       &   920.0 &   10.87 &     18.0 &    0.21 &  5.6E-18\\
\enddata
\tablecomments{The SynthIceSpec are assuming Gaussian for each absorption feature for simplicity. We assumed the center and the half of the integrated interval for the derivation of the integrated absorbance as peak position and FWHM of the absorption band, respectively.}
\end{deluxetable}

In addition to the density structure, we also compare the modeled water-ice column density with values previously reported for observations. There is no direct measurement of water ice column density for this source, but values toward various other low-mass YSOs are available for comparison. This allows the compatibility of the modeled ice compositions with the observations to be determined. Considering the correlation between water ice column density and visual extinction, which has been found for a range of visual extinctions 0 $<~A_\textrm{V}~<$ 30 mag~\citep[reviewed in][]{boogert15}, water-ice column densities on the order of $10^{19}$ cm$^{-2}$ would be expected, given the high visual extinction of Cha-MMS1. The corresponding value from the model is simply derived by integrating the number density of water ice along a pencil beam toward the core center (In observations, the water ice column density is generally measured from the optical depth of the 3 $\micron$ band relative to the dust continuum, which is expected to be confined within a very small radius for a deeply embedded object). This calculation results in the value of 1.2$\times10^{21}$ cm$^{-2}$, which is around two orders of magnitude higher than the expected observational one. This discrepancy could be caused by the model having too high a local density near the core center in particular; the $N_\textrm{H,model}$ value obtained without beam convolution is overproduced by a similar factor as the water column density, while the maximum water ice fractional abundance of $\thicksim10^{-4}$ in the model (see Figure~\ref{fig:main_rVSXi}) is just a factor of 2 higher than the measured abundances previously reported for observations~\citep{boogert13}.

Figure~\ref{fig:main_rVSXi} indicates also that the modeled dust-grain ice mantles are present to an inner radius of just a few AU, with the temperatures inside this radius being too great for ices to be maintained on the dust grains. A more extended final temperature profile would push the ices out to a region of lower density, reducing the column densities of solid-phase species. Test calculations that curtail the line-of-sight integration of ice column densities to a minimum inner radius on the order of 100 AU indeed reduces the calculated values by an appropriate factor.

These proposed influences on density and temperature could be explained by the presence of a protostellar outflow carving out the central region of the core, as long as the sightline passed through the outflow cavity before reaching the central star. Furthermore, much of the material settles in a disk even at this early stage. The observations typically don't cover the high disk columns because edge-on disks have such high columns that we don't detect the central star anymore. Such a disk and an outflow has actually been identified toward Cha-MMS1~\citep{busch20}, but is not included in the 1-D dynamical model used here.

The possibility that the chemical model itself is still overproducing the ice abundances in general cannot be completely ruled out, and the simple method of column density calculation used here could be related to this discrepancy as well. Further discussion of this topic is provided in \S~\ref{sec:discussion-water}. Here, for the purposes of further comparison of the modeled ices with observations, we make a further uniform, empirical adjustment to the modeled ice abundances, scaling all the modeled ice column densities down by a factor of 100, i.e. $f=0.01$. The water-ice column density after scaling roughly corresponds to a value that can be derived via extrapolation of Figure 7 in \citet{boogert15} to 100 mag~(1.9$\times10^{19}$ cm$^{-2}$).

\subsection{Synthesis of model ice spectra and diagnostic detectability of new ice species with JWST}~\label{sec:discussion-IR_synthesis}

Ice absorption spectra will be obtained toward the core center of Cha-MMS1 with the NIRCAM/WFSS instrument, as a part of the {\em Ice Age} program. In preparation for these observations, the near-IR spectrum expected from the core center is synthesized based on the modeled ice composition. By doing so, we are able to provide a practical diagnosis of the detectability of new ice species with JWST, as well as to get a sense of the order of magnitude of solid-phase COM abundances that would be required in order for them to be identified. A simple synthetic ice-spectrum simulation script has been developed for this purpose, with input data provided by members of the {\em Ice Age} team.
The main inputs for the simulation are the retrieved modeled ice column densities and three spectral parameters for the characterization of each absorption feature (i.e. frequency, width, and apparent band strength $A\arcmin$).

Until recently, laboratory studies for the quantitative measurement of IR spectra of interstellar ice COMs had scarcely kept pace with astronomical discoveries. Fortunately, more and more effort has been expended on this type of study in recent years, although it has typically been focused on the mid-IR regime as this is considered the fingerprint region for solid-state vibrational spectroscopy~\citep[see e.g.,][]{terwisscha18,hudson18,rachid20,hudson20a,hudson20b,terwisscha21}. For example, \citet{terwisscha18} present ice spectra (2.5$\micron$ - 20 $\micron$) for ACTD, EtOH, and DME in astronomically relevant environments, and they characterize the key absorbance features at $\lambda>5\micron$ from these reference spectra. \citet{hudson18}, \citet{hudson20a} and \citet{hudson20b} also provide direct measurements of apparent band strengths and absorption coefficients from the mid-IR spectra for amorphous ices of ACTN, DME and ACTD, respectively. From the aforementioned literature, we could directly collect the absorption feature characteristics relevant to NIRCAM/WFSS coverage (2.4$\micron$ - 5 $\micron$) for ACTN, ACTD and DME. It should be noted that ice features vary considerably depending on the chemical composition of the ice mixtures. However, there is no clear consensus on the representative ice mixture for the interstellar icy mantles, and the experimental setup for ice mixtures in the past varies significantly. For consistency, we chose the measurements from pure ice spectra as our standard.
In the case of EtOH, only the band information of absorption features at $\lambda>5\micron$ are readily available. However, the reference spectrum of pure EtOH (2.5$\micron$ to 20$\micron$) ice is publicly available on the Leiden Ice Database \footnote{https://icedb.strw.leidenuniv.nl}. This allowed us to derive the spectral parameters for the shorter wavelength region (2.5$\micron$ - 5$\micron$). The unknown apparent band strengths were estimated by using the ratio between the integrated areas of an absorption feature with different absorbance within a single IR spectrum.

In general the column density of a particular ice species, $N_\textrm{species}$, is determined according to 
\begin{equation}
    N_\textrm{species}=\textrm{ln}(10)\frac{\int_\textrm{band} I_{\lambda}}{A\arcmin_{\lambda}}
\end{equation}
where $\int_\textrm{band} I_{\lambda}$ is the integrated absorbance of the IR band under consideration and $A\arcmin_{\lambda}$ the apparent band strength. The ice column density derived from different absorption features in a single IR spectrum should correspond to each other. Thus the unknown band strengths in the near-IR region for EtOH can be constrained by using the representative absorption band for EtOH at 9.514 $\micron$ as a reference~\citep{terwisscha18}.
Table~\ref{table:ice_features} summarizes the spectral parameters used in this simulation.

Ice spectra derived from the chemodynamical models are calculated for the main simple ice constituents and the four aforementioned COMs for which IR bands are characterized. Two sets of ice column densities are tested: the regular ice composition (setup A) which is derived by directly integrating the spatial distribution of the species from the core edge to the core center,
and the ice composition scaled down by factor $f=0.01$ (as discussed in \S~\ref{sec:discussion-colden_profile}; setup B.1). Table~\ref{table:synthesized_ice_comp} shows the two sets of ice composition for the synthesis of spectra.

$^{13}$CO and $^{13}$\ce{CO2} are known as two rarer isotopologues of which ice absorption features are clearly detected in interstellar environments. As the chemical model considers the chemistry involving the main isotopologues only, the contribution of the two aforementioned species to the IR absorption features is considered by scaling the column densities of $^{12}$CO and $^{12}$CO$_{2}$ with the ratios as follow; $^{12}$\ce{CO}/$^{13}$\ce{CO}=70 and  $^{12}$\ce{CO2}/$^{13}$\ce{CO2}=86. The former is adopted from \citet{boogert02}, and the latter is derived from the solid state $^{13}$\ce{CO2}/$^{12}$\ce{CO2} ratio as a function of Galactocentric radius~\citep{boogert00}.

\begin{deluxetable}{cccc}
\tablewidth{0pt}
\tabletypesize{\footnotesize}
\tablecolumns{4}
\tablecaption{Ice compositions for the synthesis of IR spectra \label{table:synthesized_ice_comp}}
\scriptsize
\tablehead{
\colhead{} & \colhead{$N(i)_\textrm{model}$} & \multicolumn{2}{c}{0.01$N(i)_\textrm{model}$}\\
\colhead{species} & \colhead{Setup A} & \colhead{Setup B.1} & \colhead{Setup B.2}
}
\startdata
\ce{H2O}	&	1.21E+21	&	1.21E+19   &	1.21E+19	\\
CO		&	3.81E+20	&	3.81E+18	&	3.81E+18	\\
\ce{CO2}	&	4.69E+20	&	4.69E+18	&	4.69E+18	\\
\ce{CH4}	&	4.98E+19	&	4.98E+17	&	4.98E+17	\\
\ce{NH3}	&	7.36E+19	&	7.36E+17	&	7.36E+17	\\
\ce{CH3OH}	&	4.41E+19	&	4.41E+17	&	4.41E+17	\\
EtOH	&	6.05E+18	&	6.05E+16	&	1.51E+17	\\
ACTD	&	8.68E+16 &	8.68E+14	&	2.17E+15	\\
ACTN	&	2.09E+16	&	2.09E+14	&	5.23E+14	\\
DME	&	6.49E+17	&	6.49E+15	&	1.62E+16	\\
\enddata
\end{deluxetable}

\begin{figure*}
\gridline{\fig{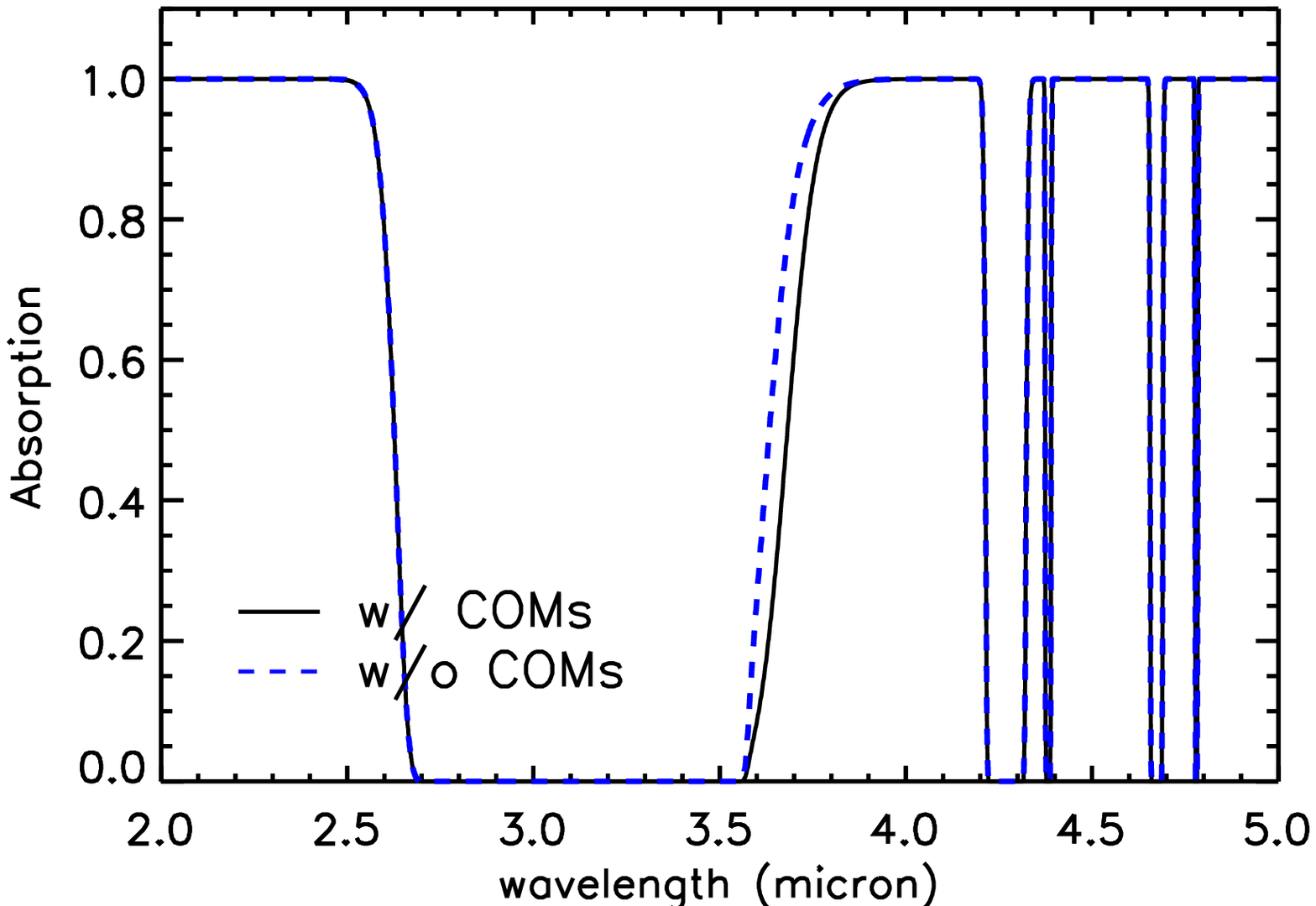}{0.5\textwidth}{}
          \fig{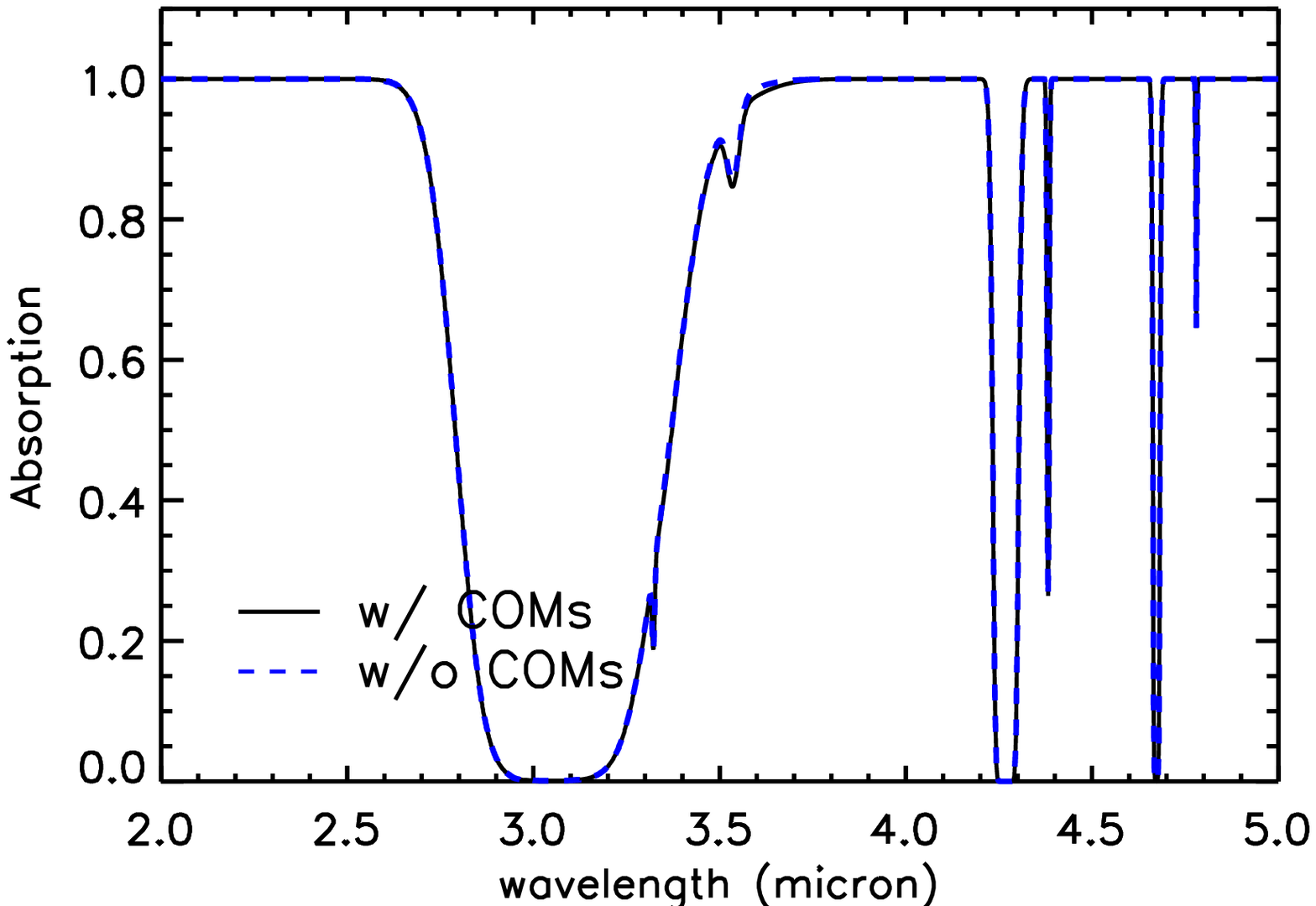}{0.5\textwidth}{}}
   \caption{Simulated IR spectrum at NIRCAM/WFSS coverage. The left panel shows the one based on the regular ice composition without scaling (setup A) while the right panel shows the one from scaled down ice composition(0.01$\times N_{i}$; setup B.1)
   \label{fig:IR_spectrum}}
\end{figure*}

\begin{figure*}
\gridline{\fig{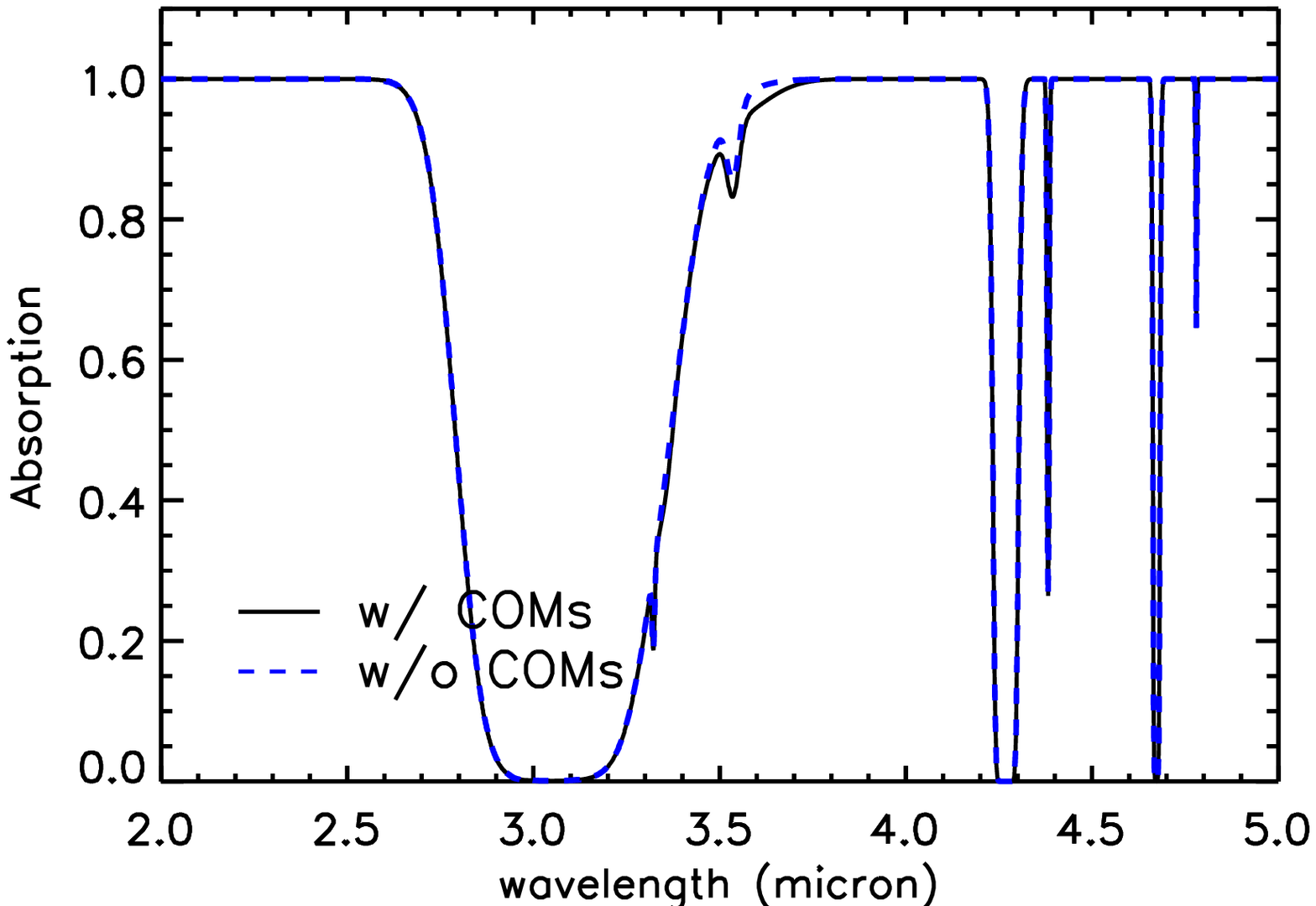}{0.5\textwidth}{}
          \fig{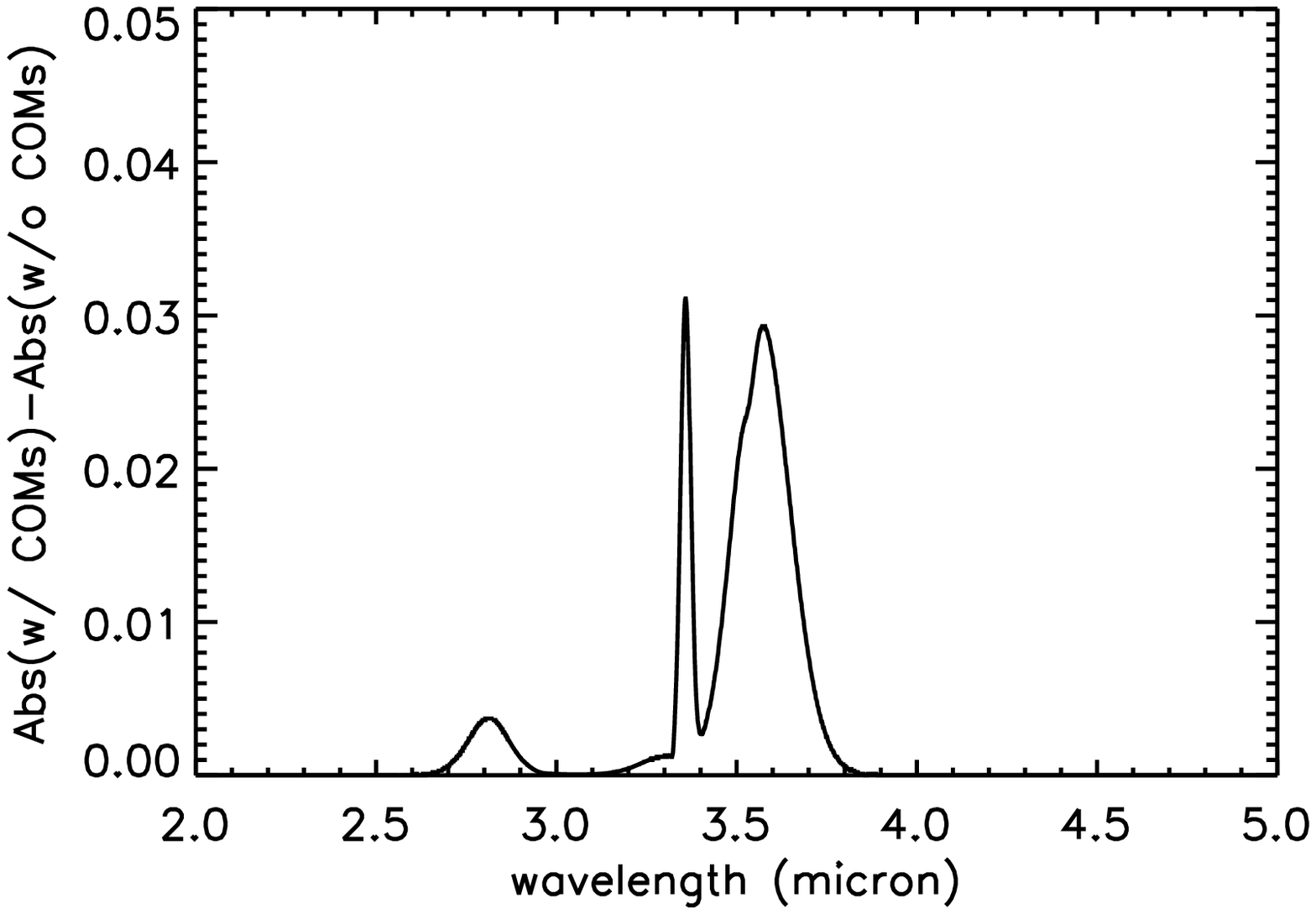}{0.5\textwidth}{}}
   \caption{Simulated IR spectrum at NIRCAM/WFSS coverage when 2.5 times higher ice abundances for COMs are assumed (setup B.2). The left panel shows the spectrum including all ice species, while the right panel has the contribution by the main ice constituents subtracted.
   \label{fig:IR_spectrum-5xCOMs_NIRCAM}}
\end{figure*}

\begin{figure*}
\gridline{\fig{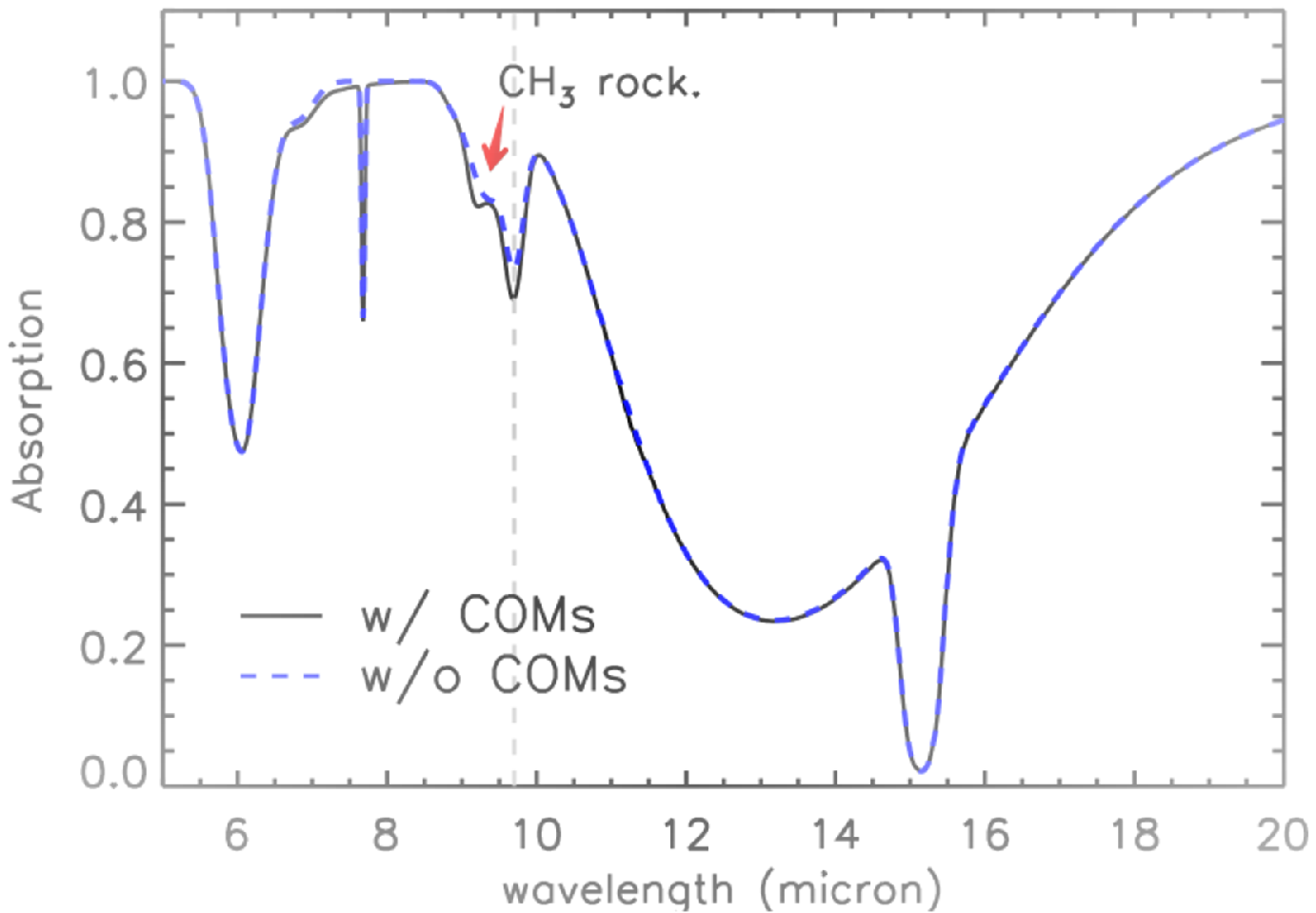}{0.5\textwidth}{}
          \fig{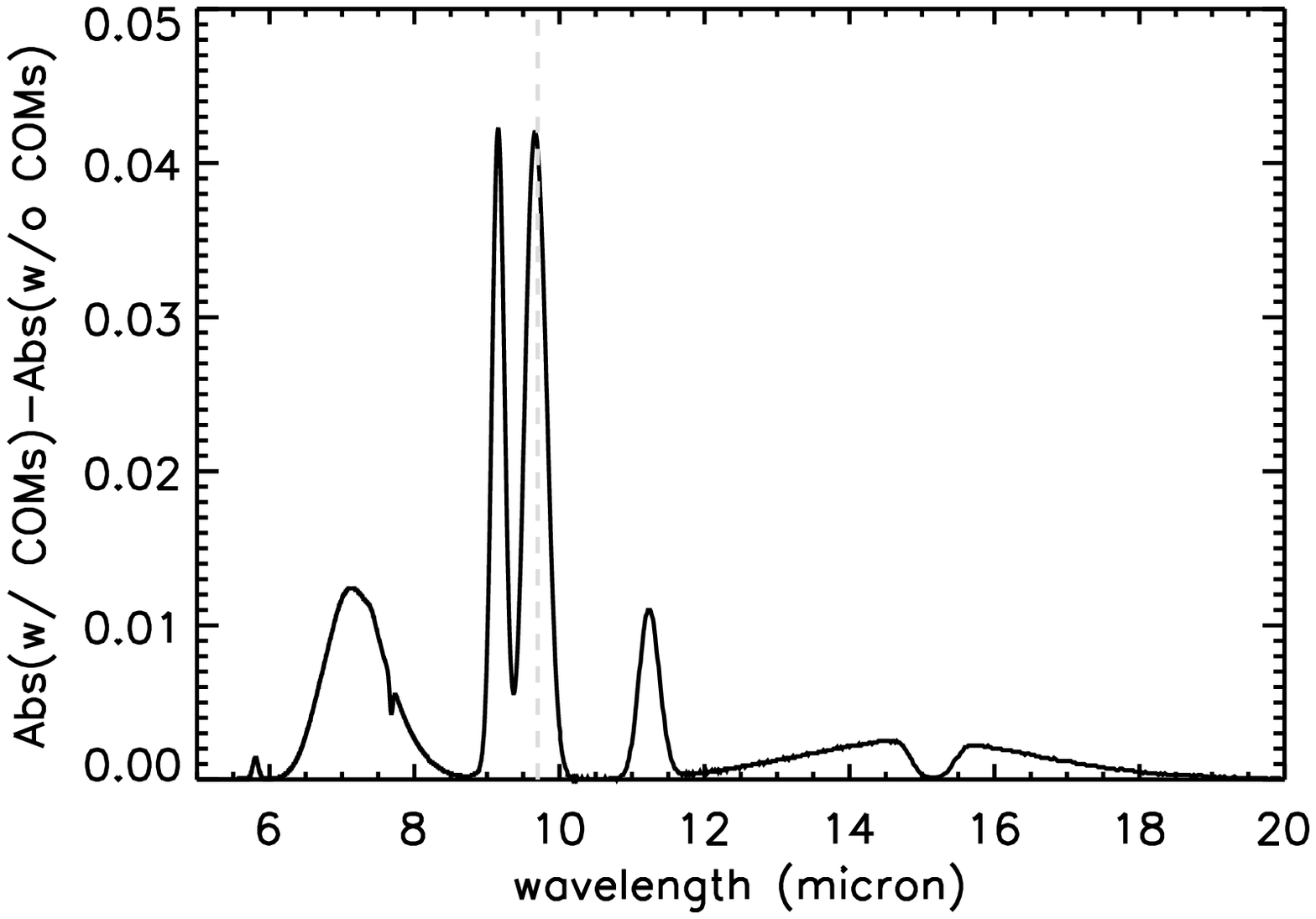}{0.5\textwidth}{}}
   \caption{Simulated IR spectrum at MIRI coverage when 2.5 times higher ice abundances for COMs are assumed (setup B.2) The left panel shows the spectrum including all ice species, while the right panel has the contribution by the main ice constituents subtracted. The grey dashed line indicates the position of 9.7$\micron$ silicate dust absorption band which is assumed to be already subtracted.
   \label{fig:IR_spectrum-5xCOMs_MIRI}}
\end{figure*}

Figure~\ref{fig:IR_spectrum} shows the IR spectra for the  NIRCAM/WFSS coverage for the two different setups of the ice composition. For each setup, a control spectrum is additionally synthesized (blue dashed line) assuming zero contribution of COMs larger than \ce{CH3OH} to the spectrum for comparison. The IR band at NIRCAM/WFSS coverage is dominated by water features and a mixture of methyl (CH$_3$) functional group vibrational stretching modes from the different ice constituents in the simulation. In particular, the spectrum of regular ice composition without scaling (setup A - left panel) is highly saturated by these features, making it hard to identify IR absorption peaks around 3.4 and 3.6 $\micron$, which are distinctively seen in setup B.1 (right panel). However, regardless of whether or not the ice composition is scaled down, the contribution of icy COMs to the IR absorption is either negligible or would not be distinguishable in the case of real observations.

To determine the approximate COM ice abundances required for them to be detected observationally, we again simulate ice spectra based on setup B.1, but this time with the ice abundances of the four COMs increased with respect to the water-ice column density by a factor of 2.5 (setup B.2). Figure~\ref{fig:IR_spectrum-5xCOMs_NIRCAM} shows the near-IR absorption spectra for setup B.2. The absorption from COM ices is more notable in this case, adding up to three absorption peaks at 2.8, 3.4 and 3.6 $\micron$ to the control IR absorption spectra. However, even if the ice abundances of EtOH are increased to as high as 1$\%$ with respect to icy \ce{H2O} in this case, the COM absorption features are still minor. Furthermore, the absorption peaks overlap with the absorption IR band of the main ice constituents. This would introduce additional uncertainty in the identification of COM features in actual observational spectra.

Although only near-IR spectra will be obtained toward the core center as part of the {\em Ice Age} project, we may extend our investigation to the mid-IR regime (5$\micron$ - 14 $\micron$; MIRI/LRS coverage) with the same ice composition to seek any distinctive ice feature from COMs; MIRI observations will be performed towards the two high-$A_\textrm{V}$ lines of sight located in the outer envelope of Cha-MMS1.

Figure~\ref{fig:IR_spectrum-5xCOMs_MIRI} shows the resulting mid-IR absorption spectra with setup B.2 (COM-rich). The IR band at 9.2 $\micron$, which is mainly attributed to the \ce{CH3} rocking mode of EtOH, could be a prospective IR band for COM detection~\citep{terwisscha18}. Although it is unclear whether such high abundances of COMs might be  achieved in typical star-forming environments, the production rates of COMs can change sensitively with UV flux. For example, in a test model with a higher extinction environment ($A_\textrm{V,bac}=5$ mag; not shown) EtOH ice abundance decreases by around a factor of four compared to the standard model. When the additional background visual extinction is not included at all ($A_\textrm{V,bac}=0$ mag; not shown), the EtOH ice abundance significantly decreases again as photodissociation overwhelms the synthesis of the species via PDI. Considering all this, the identification of COM ice might be possible with MIRI observations in an environment with a modest UV. Similar variations in COM abundances might be expected if an alternative $N_\textrm{H}$:$A_\textrm{V}$ ratio than the one used here were adopted (see also Sec.~\ref{sec:discussion-colden_profile}).

It should be noted that dust features were not considered for our calculations but only absorption features from species within ice mantles. Silicates will produce deep dust-continuum baselines around 9.7$\micron$~\citep[see L1014 IRS in figure 1;][]{oberg11}, and this could make the absorption features of COMs from 8-11 microns more difficult to detect. Here we assume that the dust continuum baseline will be constrained well and already subtracted. By observing multi lines-of-sight within a single source with the {\em Ice Age} project, we expect the dust continuum baseline would be determined with unprecedented accuracy. MIRI is expected to have at most a signal-to-noise ratio of 300. This corresponds to the detection of an absorption feature at 1\% with respect to the continuum flux at 3$\sigma$. 

\begin{deluxetable}{cccc}
\tablewidth{0pt}
\tabletypesize{\footnotesize}
\tablecolumns{4}
\tablecaption{Comparison of the abundances of gas-phase species between the model and the observation \label{table:gas_abun_comp}}
\scriptsize
\tablehead{
\colhead{species} & \colhead{observation} & \colhead{model} & \colhead{references and analytic methods} 
}
\startdata
$\ce{HC3N}$ & 1.50E-08 & 5.91E-13 & \citet[][LTE calculation]{kontinen00} \\
$\ce{CH3OH}$ & 2.27E-09 & 2.03E-09 & \citet[][LTE calculation]{kontinen00} \\
$\ce{C2S}$ & 1.03E-09 &  2.32E-11 & \citet[][LTE calculation]{kontinen00} \\
$\ce{HCN}$ & 6.67E-10 &  1.72E-09 & \citet[][Monte Carlo modeling]{tennekes06} \\
$\ce{HNC}$ & 2.00E-09 &  1.76E-09 & \citet[][Monte Carlo modeling]{tennekes06} \\
$\ce{SO}$ & 6.67E-10 &  1.90E-09 & \citet[][LTE calculation]{kontinen00} \\
$\ce{N2H+}$ & 4.67E-10 & 9.60E-11 & \citet[][LTE calculation]{kontinen00} 
\enddata
\end{deluxetable}

\subsection{Gas-phase abundances} \label{subsec:gasabun}

Gas-phase molecules have been detected in many single-dish observational studies of Cha-MMS1. For example, \citet{kontinen00} observed a selection of simple species, carbon-chain molecules, and methanol using the Swedish ESO Submillimeter Telescope (SEST), while \citet{tennekes06} determined the HCN/HNC relative abundance ratio by mapping the HCN and HNC emission with the same instrument. \citet{belloche06} measured the deuterium fractionation for molecular ions with the APEX telescope, and \citet{cordiner12} performed 7mm observation with the ATNF Mopra telescope for a selection of species including polyynes, sulphuretted carbon chains and methanol.

Although the chemical modeling presented here is focused on solid-phase abundances, the models indeed produce fractional abundance profiles for gas-phase species. In order to compare these with the observational values, column densities must be produced. Molecular column densities through the model core are produced for various offsets from the core center. These raw values are then convolved with a beam size appropriate to the observations, with the beam centered on the core.

\citet{cordiner12} provide the most extensive dataset of observed molecular lines. However, the column densities derived from our model are significantly lower than those in the Cordiner et al. dataset, by more than one order of magnitude. The large observing beam size of the Mopra telescope is attributed to this discrepancy; the observational beam size (96$\arcsec$ at 36 GHz and 77$\arcsec$ at 45 GHz) is comparable to the final radius of the outermost trajectory (r$\simeq$8660 AU). However, the two nearby sources -- Ced 110 IRS4 (class I protostar) and C2 (prestellar core) -- contaminate the observing field (see figure~\ref{fig:iceage_mapfield}), contributing to the higher molecular column densities, while the model assumes an isolated single FHSC. This may be a major factor in the poor comparison between models and observations with such a large observing beam size.

Therefore, the abundances measured with a smaller beam should be better reproduced by the chemical model. For example, the SEST observation (FWHM=55$\arcsec$ at 91GHz) shows that the abundances of some key species are roughly consistent with the modeled abundances within one order of magnitude (Table~\ref{table:gas_abun_comp}). Methanol is in fact well reproduced, and HCN and HNC are quite adequate. Sulfuretted carbon and cyanopolyyne HC$_3$N abundances are poorly reproduced; however, the overall abundance of sulfur (in whatever form) is poorly constrained in astrochemical models, as discussed by \citet{laas19}. Furthermore, HC$_3$N is traditionally considered an ``early-time'' molecule that becomes prevalent while substantial amounts of atomic carbon are still available in the gas phase;  conversely, our model specifically corresponds to a more chemically-evolved state of the gas; thus the substantial abundance of this molecule could in practice originate from a spatial scale beyond that included in our model. Ultimately, an accurate comparison of gas-phase species toward the Cha-MMS1 source may require high-resolution observations. Although there have been a few studies performing ALMA observation toward this source, it is more focused on the dynamics of the outflow, and/or quantitative measurements of the chemical composition were not fully available~\citep{busch20, allen20, maureira20}.

Using the gas-phase CO abundance reported by \citet{kontinen00}  for Cha-MMS1, \citet{cordiner12} derived an observational depletion factor of 2, assuming an undepleted CO abundance of 9.5 $\times$ $10^{-5}$~\citep{crapsi05}; this value is rather low compared with the typical CO depletion factor reported toward prestellar cores~\citep{crapsi05}. By contrast, the CO depletion factor suggested by our model is $\sim$10, assuming a convolving beam size of 50$\arcsec$ \citep[the same beam size used in the measurement of $N(\ce{H2})$ for the derivation of CO abundance;][]{tennekes06}, which is much more consistent with typical observational values for other sources. The discrepancy with the Cha-MMS1 result is likely due to the assumed beam size used in the CO observations, which is not available in the literature. The uncertainty in this value renders a comparison of observational and modeled CO depletion factors of only limited value.

\subsection{High ice column density of \ce{H2O}}~\label{sec:discussion-water}
As discussed in \S~\ref{sec:discussion-colden_profile}, the high density regions at the very center of the core may be particularly poorly represented using our simple assumption for the density structure and integration technique. The difference of ice column densities between the model and the observation could indicate the presence of an outflow cavity or other structural differences that are poorly represented by a one-dimensional dynamical treatment.

Furthermore, this divergence between the model and generally-observed values may also be related to the long-standing mystery of the oxygen budget. The water abundance (gas and ice) estimated from observations of various star-forming regions is much lower than the one expected from the known oxygen elemental abundance. This conundrum is reviewed by \citet{vandishoeck21}, who discuss two possible explanations: (i) The missing oxygen is locked up in an unidentified refractory component. If this is the case, the high abundance of water in the model is caused by a lack of gas-grain chemical network for the unidentified refractory component, which results in the oxygen budget converging to the formation of water ice. If one assumes that all volatile oxygen is locked in water, the water abundance is expected to be around 4.0$\times10^{-4}$~\citep{vandishoeck21}. This value is just a few times higher than the water-ice abundance in our model in cold environments ($\thicksim$1.0$\times10^{-4}$). (ii) Water ice abundance estimated from ice observations can be underestimated if the missing oxygen is locked up in larger $\micron$-sized grains that do not contribute to infrared ice absorption. Much observational evidence for grain growth to $\mu$m sizes has been published~\citep[e.g.][]{boogert13}; the toy model with two dust populations (0.1 $\micron$-and 10 $\micron$-sized grain) shows that a fraction $f_\textrm{large}=$ 0.01$\%$ of the grain population by number in 10 $\micron$-sized grains would catch 50$\%$ of the atomic oxygen during freeze-out and make it invisible~\citep{vandishoeck21}. However, one argument against water ice being hidden on very large grains is that after sublimation in hot cores or grain sputtering in a shocked region, the gas phase \ce{H2O} abundance along with the abundance of other O-bearing species does still not add up to the cosmic O abundance~\citep{vandishoeck21}. Possibly the 40 $\micron$ \ce{H2O} lattice mode can shed some light on this, though it lies beyond wavelength range available to JWST.

\subsection{The effect of nondiffusive chemical mechanisms} \label{subsec:non-diff}

Past astrochemical models have relied on the thermal diffusion of surface or bulk-ice radicals to drive reactions that form COMs~\citep{garrodandherbst06,garrod08a,garrod13}. However, recent detections of COMs in prestellar environments imply that COMs can be efficiently formed before star-forming regions heat up. In such colder environments, alternative mechanisms not involving thermal diffusion can become important. For example, COMs may be formed nondiffusively as the result of surface radical production in close proximity to pre-existing radicals~\citep{changandherbst16, jinandgarrod20, garrod19, garrod21}. The chemo-dynamical model used here employs the two nondiffusive mechanisms -- three-body (3-B) \citep{jinandgarrod20} and photodissociation-induced (PDI) mechanisms~\citep{garrod19}. In this section, we discuss the effect of those nondiffusive mechanisms on the composition of icy mantles by turning off either or both mechanisms.

\begin{deluxetable}{cccccc}
\tablewidth{0pt}
\tabletypesize{\footnotesize}
\tablecolumns{6}
\tablecaption{The effect of the nondiffusive mechanisms on the main ice constituents \label{table:non-diff_main}}
\scriptsize
\tablehead{
\colhead{$N(i)/N(\ce{H2O})$} & \colhead{CO} & \colhead{$\ce{CO2}$} & \colhead{$\ce{CH4}$} & \colhead{$\ce{NH3}$} & \colhead{$\ce{CH3OH}$}
}
\startdata
all on (A)	&	3.15E-01	&	3.88E-01	&	4.12E-02	&	6.09E-02	&	3.65E-02	\\
three-body off (B)	&	3.60E-01	&	2.82E-01	&	4.24E-02	&	5.62E-02	&	3.98E-02	\\
PD-induced off (C)	&	4.10E-01	&	2.66E-01	&	9.81E-02	&	6.37E-02	&	3.60E-02	\\
all off (D)  &	4.39E-01	&	1.73E-01	&	9.33E-02	&	5.89E-02	&	3.94E-02	\\
\hline
A/B	&	0.88	&	1.38	&	0.97	&	1.08	&	0.92	\\
A/C	&	0.77	&	1.46	&	0.42	&	0.96	&	1.01	\\
A/D	&	0.72	&	2.25	&	0.44	&	1.03	&	0.93	\\
\enddata
\end{deluxetable}

Table~\ref{table:non-diff_main} summarizes the results of the analysis for the main ice constituents. The relative abundances in this table are derived from the ratios of column densities, which are calculated by integrating the number density of each ice species through a line of sight toward the core center. Both nondiffusive mechanisms transfer more of the oxygen budget from CO to $\ce{CO2}$, through the reaction OH $+$ CO $\rightarrow$ $\ce{CO2}~+~$H, which results in a better reproduction of generic observations. The modeled ice abundance of \ce{CH4} significantly increases with the PDI mechanism switched off. This implies that with the PDI mechanism, a significant fraction of PD products of methane ice such as \ce{CH3} on the grain consecutively react with nearby ice species to form larger species rather than reforming \ce{CH4}, while the effect of the three-body mechanism is negligible in this case. This PDI production of methyl-group bearing molecules occurs mainly in the ice mantle, due to the much larger quantities of material stored beneath the ice surface. The solid-phase abundances of \ce{CH3OH} and \ce{NH3} barely change regardless of the choice of nondiffusive mechanism, reproducing the observed abundances well. Aside from the above-mentioned effects, the influence of nondiffusive mechanisms on the main, simple ice constituents is very minor.

\begin{deluxetable}{ccccccccccc}
\tablewidth{0pt}
\tabletypesize{\footnotesize}
\tablecolumns{11}
\tablecaption{The effect of the nondiffusive mechanisms on the O-bearing ice species \label{table:non-diff_o-spec}}
\scriptsize
\tablehead{
\colhead{$N(i)/N(\ce{H2O})$} & \colhead{EtOH} & \colhead{FA} &  \colhead{ACTD} & \colhead{MF} & \colhead{DME} & \colhead{ACTN} & \colhead{GA} &  \colhead{AA} & \colhead{EG} & \colhead{MM}
}
\startdata
all on (A)	&	5.01E-03	&	2.10E-04	&	7.18E-05	&	8.31E-05	&	5.37E-04	&	1.73E-05	&	2.73E-05	&	1.01E-04	&	2.92E-05	&	8.38E-05	\\
three-body off (B)	&	5.21E-03	&	1.43E-04	&	4.49E-05	&	6.67E-06	&	4.44E-04	&	1.42E-05	&	4.57E-06	&	2.65E-05	&	5.88E-06	&	1.13E-05	\\
PD-induced off (C)	&	6.14E-04	&	2.89E-04	&	2.11E-04	&	6.88E-05	&	2.23E-04	&	1.26E-05	&	3.54E-06	&	3.07E-05	&	7.64E-06	&	9.17E-06	\\
all off (D)  &	2.03E-04	&	1.14E-04	&	2.02E-05	&	3.01E-08	&	2.15E-04	&	5.02E-06	&	5.87E-13	&	4.28E-10	&	4.34E-13	&	1.05E-11	\\
\hline
A/B	&	0.96	&	1.47	&	1.60	&	12.47	&	1.21	&	1.22	&	5.97	&	3.80	&	4.97	&	7.42	\\
A/C	&	8.15	&	0.73	&	0.34	&	1.21	&	2.41	&	1.38	&	7.70	&	3.27	&	3.83	&	9.14	\\
A/D	&	24.70	&	1.85	&	3.55	&	$>$ 1000	&	2.49	&	3.45	&	$>$ 1000	&	$>$ 1000	&	$>$ 1000	&	$>$ 1000	\\
\enddata
\end{deluxetable}

Table~\ref{table:non-diff_o-spec} shows the results for solid-phase O-bearing COMs. As seen in the last row of this table, the inclusion of the nondiffusive mechanisms increases the abundances of COMs in general, as it provides additional pathways to form large compounds without a temperature dependence. In particular, the abundances of the three structural isomers of \ce{C2H4O2} (MF, GA, and AA) and two \ce{CH2OH}-bearing COMs (EG and MM) increase by more than three orders of magnitudes with the inclusion of any of the nondiffusive mechanisms. It is notable that the change in chemical behavior produced by such changes can be non-linear; when one strong mechanism is removed, a weaker one may take over to replace (partially) what would otherwise be done by the stronger mechanism. 
Simple comparison of these models therefore provides only a limited indication of which mechanism is most influential in the development of chemical complexity in general, as the chemical behavior of the species in the various nondiffusive setups is different.
For example, MF is more efficiently formed via 3-B excited formation \citep[see][for further detail]{jinandgarrod20} than via the PDI mechanism, although either mechanism will promote the formation of this species. The ice abundance of EtOH is higher when only the PDI mechanism switched on rather than the both, while FA and ACTD are the opposite cases; the ice abundances of the two COMs decrease with the PDI mechanism switched on.
This is because \ce{OH} and \ce{CH2} radicals -- which are the photodissociation products of \ce{H2O} and \ce{CH4}, respectively, at early times in the model evolution -- are consumed in various nondiffusive reactions, to form larger species. The main formation routes for FA and ACTD involve \ce{OH} and \ce{CH2} radicals, respectively: FA is synthesized either by the reaction $\ce{CO} + \ce{OH} \rightarrow \ce{COOH}$, followed by hydrogenation, or by the addition of OH and HCO radicals. The two-stage reaction between \ce{CH2} and \ce{H2CO}, in which \ce{CH2} abstracts an H-atom from formaldehyde with the product radicals then immediately recombining, is one of the main formation routes for \ce{CH3CHO}. Thus, the PDI mechanism is critical to transfer the CNO budget to various species at early times, developing the chemical complexity of the bulk ice.

\subsection{Implications of this study and future perspectives}
Motivated by the imminent observational improvements to be provided by JWST, laboratory experiments have recently been undertaken to characterize the spectroscopic behavior of larger species in the ice~\citep{terwisscha18, rachid20, hudson20a, hudson20b, terwisscha21}. As these quantitative measurements have accumulated, the demand has increased for the application of the experimental outputs into a chemical model; this combined method can more practically probe the strategy needed to identify new ice species with JWST. As ice spectra are influenced by the ice composition as well as the physical structure of the solid (ices and dust grains), it is important to obtain a plausible estimate for the ice composition with a chemical model. This initial study begins to meet this demand. However, our results show that even if ice abundances of COMs are predicted to be as high as 0.5\% with respect to water ice, the absorption features of these species are not distinguishable from the water absorption features. This is a good example highlighting the importance of a comprehensive study: the detectability of the ice species can be securely diagnosed with careful consideration of spectroscopic characteristics as well as molecular content before observations follow.

Based on the combination of outstanding features such as a relatively strong IR band, high ice abundance, and the potential to be enhanced in high UV-flux environments, this study indicates EtOH as a good candidate for a new ice detection with MIRI observations, given an appropriate ice abundance. So far, EtOH has not been securely identified in ice observations, but this species is one of the three COMs (MF, ACTD, and EtOH) that is considered to be a tentatively identified species~\citep[e.g.][]{terwisscha18, terwisscha21}, which supports the finding of this study. Only gas-phase detections of EtOH in hot core regions has been reported in the Milky Way, indirectly implying a possible origin for this species in the sublimated ices. However, if EtOH were to be detected in cold gas at a cloud edge, perhaps as the result of photodesorption from the icy grain surfaces, this would separately indicate the richness of EtOH in the ice. Such an observational probe could be a reasonable diagnostic to find candidate regions for new EtOH ice detection with JWST in the future.

The results of this study show that the detection of new ice species may be demanding even with JWST. However, it is noteworthy that the IR spectroscopic considerations in this study were only possible for a handful of COMs, due to a lack of quantitative measurements of IR data. Spectroscopic behavior can be significantly different between species, and some molecules that are not covered in this study yet could be ``IR-bright" (strong IR absorption band). For example, a recent study found that the infrared absorbance of propynal's alkyne bond is about 30,000\% stronger than the corresponding feature in acetylene~\citep{hudsonandgerakines19}. Although propynal's ice abundance is not high enough in our FHSC chemical model, further IR investigation of other larger interstellar species may unveil new IR bright COMs that are also abundant in the chemical model. With the interplay between laboratory work and chemical modeling studies, physical environments where such IR-bright COMs are efficiently synthesized can be also examined. This comprehensive study provides firm ground to further explore new ice species in the future. 

 \section{Conclusions}

To timely diagnose the detectability of new ice species with JWST, we predict the chemical compositions within the dust-grain ice mantles in the young star-forming core Cha-MMS1 using a 1-D chemo-dynamical model. The expected IR spectrum based on the given ice composition is simulated. The main conclusions of this study are enumerated below:

\begin{enumerate}
\item In the chemo-dynamical model, the relative abundances of the main ice constituents with respect to water toward the core center match well with generic observational values, providing a firm basis to further explore the ice chemistry.

\item The model predicts relatively high solid-phase abundances (i.e. $>0.01$\% with respect to water ice column density) of seven large oxygen-bearing species (ethanol, acetaldehyde, methyl formate, dimethyl ether, formic acid, acetic acid, and methoxy methanol). All but acetic acid and methoxy methanol have, either directly or indirectly, implied their presence within the ice mantles through the past observations.

\item For the large molecules, the abundance profiles in the model are found to depend on which functional groups are present; -\ce{CH3} and -\ce{CH2OH} might be the key functional groups delineating the spatial distributions of species.

\item The IR spectrum is synthesized based on the modeled ice column densities of four COMs where IR band information is known, as well as the main ice constituents. The contribution of COMs to IR absorption is minor compared to the main ice constituents, making the identification of COMs with the NIRCAM/WFSS instrument unlikely.

\item Mid-IR observations of COM-rich (ice abundances of COMs w.r.t. the water ice column density $>1$\%) environments could reveal distinctive ice features of COMs. As this ice feature overlaps with silicate features from dust grains at around 9.7$\micron$, accurate treatment for the continuum subtraction is critical.

\item The results of this study show that the detection of new icy COMs with JWST could still be challenging. However, Further IR investigation of other larger interstellar species may unveil new IR bright COMs that are also abundant enough within the ice mantles to be identified.

\end{enumerate}

\section*{Acknowledgements}

We thank V. Wakelam and P. Caselli for helpful discussions about initial elemental abundances and the calculation of water ice column densities. We thank P. Gratier and A. Taillard for developing and providing an early version of their synthetic ice spectrum simulator within the IceAge consortium, which informed the code used in this study. We also gratefully acknowledge the input spectral data provided by E. Dartois and V. Wakelam for use with the simulator. MJ and RTG thank the Space Telescope Science Institute for funding through grant JWST-ERS-01309.020-A. Chemical model development was funded by the National Science Foundation through the Astronomy \& Astrophysics program (grant No. AST 19-06489). Development of the coupled chemical and dynamical modeling treatment was funded by the NASA Astrophysics Theory program (grant number 80NSSC18K0558) and is supported in part by NASA NSSC20K0533, NSF AST-1815784, and NSF AST-1716259.


\begin{thebibliography}{}

\bibitem[Aikawa et al.(2008)]{aikawa08} Aikawa, Y., Wakelam, V., Garrod, R.~T., et al.\ 2008, \apj, 674, 984. doi:10.1086/524096

\bibitem[Allen et al.(2020)]{allen20} Allen, V., Cordiner, M.~A., Adande, G., et al.\ 2020, arXiv:2010.01151

\bibitem[Arce et al.(2008)]{arce08} Arce H.~G., Santiago-Garcia J., J{\o}rgensen J.~K., Tafalla M., \& Bachiller R.\ 2008, \apjl, 681, L21

\bibitem[Bacmann et al.(2012)]{bacmann12} Bacmann A., Taquet V., Faure A., Kahane C., \& Ceccarelli C.\ 2012, \aap, 541, L21

\bibitem[Balucani et al.(2015)]{balucani15} Balucani, N., Ceccarelli, C., \& Taquet, V.\ 2015, \mnras, 449, L16. doi:10.1093/mnrasl/slv009

\bibitem[Baluch et al.(1992)]{baluch92} Baulch, D.~L., Cobos, C.~J., Cox, R.A, et al.\ 1992, J. Phys. Chem. Ref. Data, 21, 411

\bibitem[Barger et al.(2021)]{Barger21} Barger, C.~J., Lam, K.~H., Garrod, R.~T., Li, Z.~Y., Davis, S.~W. \& Herbst, E. \ 2021, \aap, {\em accepted}

\bibitem[Belloche et al.(2006)]{belloche06} Belloche, A., Parise, B., van der Tak, F.~F.~S., et al.\ 2006, \aap, 454, L51

\bibitem[Belloche et al.(2011)]{belloche11} Belloche, A., Parise, B., Schuller, F., et al.\ 2011, \aap, 535, A2. doi:10.1051/0004-6361/201117276

\bibitem[Bergner et al.(2019)]{bergner19} Bergner, J.~B., {\"O}berg, K.~I., \& Rajappan, M.\ 2019, \apj, 874, 115. doi:10.3847/1538-4357/ab07b2

\bibitem[Bernstein et al.(2002)]{bernstein02} Bernstein, M.~P., Dworkin, J.~P., Sandford, S.~A., et al.\ 2002, \nat, 416, 401. doi:10.1038/416401a

\bibitem[Boogert et al.(2000)]{boogert00} Boogert, A.~C.~A., Ehrenfreund, P., Gerakines, P.~A., et al.\ 2000, \aap, 353, 349

\bibitem[Boogert et al.(2002)]{boogert02} Boogert, A.~C.~A., Blake, G.~A., \& Tielens, A.~G.~G.~M.\ 2002, \apj, 577, 271. doi:10.1086/342176

\bibitem[Boogert et al.(2008)]{boogert08} Boogert, A.~C.~A., Pontoppidan, K.~M., Knez, C., et al.\ 2008, \apj, 678, 985. doi:10.1086/533425

\bibitem[Boogert et al.(2013)]{boogert13} Boogert, A.~C.~A., Chiar, J.~E., Knez, C., et al.\ 2013, \apj, 777, 73. doi:10.1088/0004-637X/777/1/73

\bibitem[Boogert et al.(2015)]{boogert15} Boogert, A.~C.~A., Gerakines, P.~A., \& Whittet, D.~C.~B.\ 2015, \araa, 53, 541

\bibitem[Bottinelli et al.(2010)]{bottinelli10} Bottinelli S., Boogert A.~C.~A., Bouwman J. et al.\ 2010, \apj, 718, 1100

\bibitem[Blake et al.(1987)]{blake87} Blake G.~A., Sutton E.~C., Masson C.~R., \& Phillips T.~G.\ 1987, \apj, 315, 621

\bibitem[Bohlin et al.(1978)]{bohlin78} Bohlin, R.~C., Savage, B.~D., \& Drake, J.~F.\ 1978, \apj, 224, 132. doi:10.1086/156357

\bibitem[Busch et al.(2020)]{busch20} Busch, L.~A., Belloche, A., Cabrit, S., et al.\ 2020, \aap, 633, A126

\bibitem[Cernicharo et al.(2012)]{cernicharo12} 
Cernicharo, J., Marcelino, N., Roueff, E., et al.
2012, \apjl, 759, L43 

\bibitem[Chang \& Herbst(2016)]{changandherbst16} 
Chang, Q., \& Herbst, E.
2016, \apj, 819, 145 

\bibitem[Chang et al.(2020)]{chang20} Chang, P., Davis, S.~W., \& Jiang, Y.-F.\ 2020, \mnras, 493, 5397. doi:10.1093/mnras/staa573

\bibitem[Chapman et al.(2009)]{chapman09} Chapman, N.~L., Mundy, L.~G., Lai, S.-P., et al.\ 2009, \apj, 690, 496. doi:10.1088/0004-637X/690/1/496

\bibitem[Chuang et al.(2016)]{chuang2016} 
Chuang, K.-J., Fedoseev, G., Ioppolo, S., et al.\ 2016, \mnras, 455, 1702

\bibitem[Cordiner et al.(2012)]{cordiner12} Cordiner, M.~A., Charnley, S.~B., Wirstr{\"o}m, E.~S., et al.\ 2012, \apj, 744, 131. doi:10.1088/0004-637X/744/2/131

\bibitem[Crapsi et al.(2005)]{crapsi05} Crapsi, A., Caselli, P., Walmsley, C.~M., et al.\ 2005, \apj, 619, 379. doi:10.1086/426472

\bibitem[Davis \& Gammie(2020)]{davisandgammie20} Davis, S.~W. \& Gammie, C.~F.\ 2020, \apj, 888, 94. doi:10.3847/1538-4357/ab5950

\bibitem[Dzib et al.(2018)]{dzib18} Dzib, S.~A., Loinard, L., Ortiz-Le{\'o}n, G.~N., et al.\ 2018, \apj, 867, 151

\bibitem[Fayolle et al.(2015)]{fayolle15} Fayolle E.~C., {\"O}berg K. I., Garrod R.~T., van Dishoeck E.~F., \& Bisschop S.~E.\ 2015, \aap, 576, A45

\bibitem[Fedoseev et al.(2015)]{fedoseev15} 
Fedoseev, G., Cuppen, H.~M., Ioppolo, S., et al.\ 2015, Monthly Notices of the Royal Astronomical Society, 448, 1288

\bibitem[Fedoseev et al.(2017)]{fedoseev17}
Fedoseev, G., Chuang, K.-J., Ioppolo, S., et al.\ 2017, The Astrophysical Journal, 842, 52

\bibitem[Furuya et al.(2012)]{furuya12} Furuya, K., Aikawa, Y., Tomida, K., et al.\ 2012, \apj, 758, 86. doi:10.1088/0004-637X/758/2/86

\bibitem[Garrod \& Herbst(2006)]{garrodandherbst06}
Garrod R.~T., \& Herbst E. 
2006, \aap, 457, 927

\bibitem[Garrod et al.(2008)]{garrod08a} Garrod, R.~T., Widicus Weaver, S.~L., \& Herbst, E. 2008, \apj, 682, 283 

\bibitem[Garrod(2008)]{garrod08b} Garrod, R.~T.\ 2008, Astronomy and Astrophysics, 491, 239

\bibitem[Garrod \& Pauly(2011)]{garrodandpauly11} Garrod, R.~T. \& Pauly, T.\ 2011, \apj, 735, 15

\bibitem[Garrod(2013)]{garrod13} Garrod, R.~T.\ 2013, \apj, 765, 60

\bibitem[Garrod(2019)]{garrod19} Garrod, R.~T.\ 2019, \apj, 884, 69. doi:10.3847/1538-4357/ab418e

\bibitem[Garrod et al.(2022)]{garrod21} Garrod, R.~T., Jin, M., Matis, K.~A., et al.\ 2022, \apjs, 259, 1. doi:10.3847/1538-4365/ac3131

\bibitem[Gerakines et al.(1996)]{gerakines96} Gerakines, P.~A., Schutte, W.~A., \& Ehrenfreund, P.\ 1996, \aap, 312, 289

\bibitem[Gerakines \& Hudson(2020)]{gerakines20} Gerakines, P.~A. \& Hudson, R.~L.\ 2020, \apj, 901, 52. doi:10.3847/1538-4357/abad39

\bibitem[Gerin et al.(2015)]{gerin15} Gerin, M., Pety, J., Fuente, A., et al.\ 2015, \aap, 577, L2. doi:10.1051/0004-6361/201525777

\bibitem[Goldsmith(2001)]{goldsmith01} Goldsmith, P.~F.\ 2001, \apj, 557, 736. doi:10.1086/322255

\bibitem[Golden(1958)]{golden58} Golden, S.\ 1958, J. Chem. Phys. 29, 61.

\bibitem[Graedel et al.(1982)]{graedel82} Graedel, T.~E., Langer, W.~D., \& Frerking, M.~A.\ 1982, \apjs, 48, 321

\bibitem[Hamberg et al.(2010)]{hamberg10} Hamberg, M., {\"O}sterdahl, F., Thomas, R.~D., et al.\ 2010, \aap, 514, A83. doi:10.1051/0004-6361/200913891

\bibitem[Herbst \& van Dishoeck(2009)]{HvD2009} Herbst, E. \& van Dishoeck, E.~F.\ 2009, \araa, 47, 427. doi:10.1146/annurev-astro-082708-101654

\bibitem[Hincelin et al.(2011)]{hincelin11} Hincelin, U., Wakelam, V., Hersant, F., et al.\ 2011, \aap, 530, A61

\bibitem[Hincelin et al.(2013)]{hincelin13} Hincelin, U., Wakelam, V., Commer{\c{c}}on, B., et al.\ 2013, \apj, 775, 44. doi:10.1088/0004-637X/775/1/44

\bibitem[Hincelin et al.(2016)]{hincelin16} Hincelin, U., Commer{\c{c}}on, B., Wakelam, V., et al.\ 2016, \apj, 822, 12. doi:10.3847/0004-637X/822/1/12

\bibitem[Hudson et al.(2018)]{hudson18} Hudson, R.~L. Gerakines, P.~A. \& Ferrante, R.~F.\ 2018, Spectrochimica Acta Part A: Molecular and Biomolecular Spectroscopy, 193, 33. doi:10.1016/j.saa.2017.11.055 

\bibitem[Hudson \& Ferrante(2020)]{hudson20a} Hudson, R.~L. \& Ferrante, R.~F.\ 2020, \mnras, 492, 283. doi:10.1093/mnras/stz3323

\bibitem[Hudson \& Gerakines(2019)]{hudsonandgerakines19} Hudson, R.~L. \& Gerakines, P.~A.\ 2019, \mnras, 482, 4009. doi:10.1093/mnras/sty2821

\bibitem[Hudson et al.(2020)]{hudson20b} Hudson, R.~L., Yarnall, Y.~Y. \& Coleman, F.~M.\ 2020, Spectrochimica Acta Part A: Molecular and Biomolecular Spectroscopy, 233, 118217. doi:10.1016/j.saa.2020.118217

\bibitem[Ioppolo et al.(2020)]{Ioppolo20} Ioppolo, S., Fedoseev, G., Chuang, K.-J., et al.\ 2020, Nature Astronomy. doi:10.1038/s41550-020-01249-0

\bibitem[Jackson \& Montroll(1958)]{jacksonandmontroll58} Jackson, J.~L., Montroll, E.~W.\ 1958, J. Chem. Phys. 28, 1101

\bibitem[Jackson(1959a)]{jackson59a} Jackson, J.~L., \ 1959, J. Chem. Phys. 31, 154

\bibitem[Jackson(1959b)]{jackson59b} Jackson, J.~L., \ 1959, J. Chem. Phys. 31, 722

\bibitem[Jenkins(2009)]{jenkins09} Jenkins, E.~B.\ 2009, \apj, 700, 1299

\bibitem[Jiang et al.(2019)]{jiang19} Jiang, Y.-F., Blaes, O., Stone, J.~M., et al.\ 2019, \apj, 885, 144. doi:10.3847/1538-4357/ab4a00

\bibitem[Jin et al.(2015)]{jin15} Jin, M., Lee, J.-E., \& Kim, K.-T.\ 2015, \apjs, 219, 2

\bibitem[Jin \& Garrod(2020)]{jinandgarrod20} Jin, M. \& Garrod, R.~T.\ 2020, \apjs, 249, 26

\bibitem[J{\o}rgensen et al.(2012)]{jorgensen12} J{\o}rgensen, J.~K., Favre, C., Bisschop, S.~E., et al.\ 2012, \apjl, 757, L4

\bibitem[J{\o}rgensen et al.(2016)]{jorgensen16} J{\o}rgensen, J.~K., van der Wiel, M.~H.~D., Coutens, A., et al.\ 2016, \aap, 595, A117

\bibitem[J{\o}rgensen, Belloche \& Garrod(2020)]{jorgensen20} J{\o}rgensen, J.~K., Belloche, A., \& Garrod, R.~T.\ 2020, \araa, 58, 727. doi:10.1146/annurev-astro-032620-021927

\bibitem[Kalv{\={a}}ns(2018)]{Kalvans18} Kalv{\={a}}ns, J.\ 2018, \mnras, 478, 2753. doi:10.1093/mnras/sty1172

\bibitem[Kontinen et al.(2000)]{kontinen00} Kontinen, S., Harju, J., Heikkil{\"a}, A., et al.\ 2000, \aap, 361, 704

\bibitem[Kuiper et al.(2010)]{kuiper10} Kuiper, R., Klahr, H., Dullemond, C., et al.\ 2010, \aap, 511, A81. doi:10.1051/0004-6361/200912355

\bibitem[Laas \& Caselli(2019)]{laas19} Laas, J.~C. \& Caselli, P.\ 2019, \aap, 624, A108. doi:10.1051/0004-6361/201834446

\bibitem[Larson(1969)]{larson69} Larson, R.~B.\ 1969, \mnras, 145, 271. doi:10.1093/mnras/145.3.271

\bibitem[Lee et al. (1978)]{lee78} Lee, J.~H., Michael, J.~V., Payne, W.~A., \& Stief, L.~J., J. Chem. Phys., 68

\bibitem[Lehtinen et al.(2003)]{lehtinen03} Lehtinen, K., Harju, J., Kontinen, S., et al.\ 2003, \aap, 401, 1017. doi:10.1051/0004-6361:20030185

\bibitem[Majewski et al.(2011)]{majewski11} Majewski, S.~R., Zasowski, G., \& Nidever, D.~L.\ 2011, \apj, 739, 25. doi:10.1088/0004-637X/739/1/25

\bibitem[Masunaga et al.(1998)]{masunaga98} Masunaga, H., Miyama, S.~M., \& Inutsuka, S.-.I \ 1998, \apj, 495, 346. 
\bibitem[Maureira et al.(2020)]{maureira20} Maureira, M.~J., Arce, H.~G., Dunham, M.~M., et al.\ 2020, \mnras, doi:10.1093/mnras/staa2894

\bibitem[Mart{\'\i}n-Dom{\'e}nech et al.(2020)]{martin-domenech20} Mart{\'\i}n-Dom{\'e}nech, R., {\"O}berg, K.~I., \& Rajappan, M.\ 2020, \apj, 894, 98. doi:10.3847/1538-4357/ab84e8

\bibitem[McClure(2009)]{mcclure09} McClure, M.\ 2009, \apjl, 693, L81. doi:10.1088/0004-637X/693/2/L81

\bibitem[McGuire(2018)]{Mcguire18} McGuire, B.~A.\ 2018, \apjs, 239, 17. doi:10.3847/1538-4365/aae5d2

\bibitem[Minissale et al.(2016)]{minissale16} Minissale, M., Congiu, E., \& Dulieu, F.\ 2016, \aap, 585, A146

\bibitem[Nejad et al.(1990)]{nejad90} Nejad, L. A. M., Williams, D. A., \& Charnley, S. B. 1990, ApJ, 246, 183 

\bibitem[Neufeld et al.(2015)]{neufeld15} Neufeld, D.~A., Indriolo, N., DeWitt, C.~N., et al.\ 2015, IAU General Assembly

\bibitem[{\"O}berg et al.(2010)]{oberg10} {\"O}berg K. I., Bottinelli S., J{\o}rgensen J. K., \& van Dishoeck E. F.\ 2010, \apj, 716, 825
 
\bibitem[{\"O}berg et al.(2011)]{oberg11} {\"O}berg, K.~I., Boogert, A.~C.~A., Pontoppidan, K.~M., et al.\ 2011, \apj, 740, 109. doi:10.1088/0004-637X/740/2/109
 
\bibitem[Persi et al.(2001)]{persi01} Persi, P., Marenzi, A.~R., G{\'o}mez, M., et al.\ 2001, \aap, 376, 907. doi:10.1051/0004-6361:20010962
 
\bibitem[Penston(1969)]{penston69} Penston, M.~V.\ 1969, \mnras, 144, 425. doi:10.1093/mnras/144.4.425
 
\bibitem[Rachid et al.(2020)]{rachid20} Rachid, M.~G., Terwisscha van Scheltinga, J., Koletzki, D., et al.\ 2020, \aap, 639, A4. doi:10.1051/0004-6361/202037497

\bibitem[Reipurth et al.(1996)]{reipurth96} Reipurth, B., Nyman, L.-A., \& Chini, R.\ 1996, \aap, 314, 258

\bibitem[Rodgers \& Charnley(2003)]{rodgersandcharnley03} Rodgers, S.~D. \& Charnley, S.~B.\ 2003, \apj, 585, 355. 

\bibitem[Schlafly et al.(2016)]{schlafly16} Schlafly, E.~F., Meisner, A.~M., Stutz, A.~M., et al.\ 2016, \apj, 821, 78. doi:10.3847/0004-637X/821/2/78

\bibitem[Schutte et al.(1999)]{schutte99} Schutte, W.~A., Boogert, A.~C.~A., Tielens, A.~G.~G.~M., et al.\ 1999, \aap, 343, 966

\bibitem[Scibelli \& Shirley(2020)]{scibelli20} Scibelli, S. \& Shirley, Y.\ 2020, \apj, 891, 73. doi:10.3847/1538-4357/ab7375

\bibitem[Shannon et al.(2013)]{shannon13} Shannon, R., Blitz, M., Goddard, A. et al. 2013, Nature Chem 5, 745 doi: 10.1038/nchem.1692

\bibitem[Shingledecker et al.(2018)]{shingledecker18} Shingledecker, C.~N., Tennis, J., Le Gal, R., et al.\ 2018, \apj, 861, 20. doi:10.3847/1538-4357/aac5ee

\bibitem[Slate et al.(2020)]{slate20} Slate, E.~C.~S., Barker, R., Euesden, R.~T., et al.\ 2020, \mnras, 497, 5413. doi:10.1093/mnras/staa2436

\bibitem[Stone et al.(2020)]{stone20} Stone, J.~M., Tomida, K., White, C.~J., et al.\ 2020, \apjs, 249, 4. doi:10.3847/1538-4365/ab929b

\bibitem[Taquet et al.(2014)]{taquet14} Taquet, V., Charnley, S.~B., \& Sipil{\"a}, O.\ 2014, \apj, 791, 1.

\bibitem[Tennekes et al.(2006)]{tennekes06} Tennekes, P.~P., Harju, J., Juvela, M., et al.\ 2006, \aap, 456, 1037. doi:10.1051/0004-6361:20040294

\bibitem[Tercero et al.(2018)]{tercero18} Tercero, B., Cuadrado, S., L{\'o}pez, A., et al.\ 2018, \aap, 620, L6. doi:10.1051/0004-6361/201834417

\bibitem[Terwisscha van Scheltinga et al.(2018)]{terwisscha18} Terwisscha van Scheltinga, J., Ligterink, N.~F.~W., Boogert, A.~C.~A., et al.\ 2018, \aap, 611, A35. doi:10.1051/0004-6361/201731998

\bibitem[Terwisscha van Scheltinga et al.(2021)]{terwisscha21} Terwisscha van Scheltinga, J., Marcandalli, G., McClure, M.~K., et al.\ 2021, \aap, 651, A95. doi:10.1051/0004-6361/202140723

\bibitem[van Dishoeck et al.(2021)]{vandishoeck21} van Dishoeck, E.~F., Kristensen, L.~E., Mottram, J.~C., et al.\ 2021, \aap, 648, A24. doi:10.1051/0004-6361/202039084

\bibitem[Vastel et al.(2014)]{vastel14} Vastel, C., Ceccarelli, C., Lefloch, B., \& Bachiller, R. 2014, \apjl, 795, L2 

\bibitem[Vasyunin \& Herbst(2013)]{vasyuninandherbst13} Vasyunin, A.~I. \& Herbst, E.\ 2013, \apj, 769, 34. doi:10.1088/0004-637X/769/1/34

\bibitem[Wakelam \& Herbst(2008)]{wakelamandherbst08} Wakelam, V. \& Herbst, E.\ 2008, \apj, 680, 371

\bibitem[Willis et al.(2020)]{willis20} Willis, E.~R., Garrod, R.~T., Belloche, A., et al.\ 2020, \aap, 636, A29

\bibitem[Xue et al.(2016)]{xue16} Xue, M., Jiang, B.~W., Gao, J., et al.\ 2016, \apjs, 224, 23. doi:10.3847/0067-0049/224/2/23

\end{thebibliography}
\end{document}